\documentclass{article}

\usepackage{graphicx}
\usepackage{verbatim}
\usepackage{tikz}
\usepackage{pgfplots}
\usepackage{pgfplotstable}
\usepackage{amsmath}
\usepackage{caption}
\usepackage{subcaption}
\usepackage{bm}

\newcommand\be{\begin{equation}}
\newcommand\ee{\end{equation}}
\newcommand\bea{\begin{eqnarray}}
\newcommand\eea{\end{eqnarray}}

\newcommand\base{{\rm base}}

\newcommand{\free}{{\rm outer}}
\newcommand{\tw}{{\tilde{w}}}

\makeatletter
\newcommand{\pushright}[1]{\ifmeasuring@#1\else\omit\hfill$\displaystyle#1$\fi\ignorespaces}
\newcommand{\pushleft}[1]{\ifmeasuring@#1\else\omit$\displaystyle#1$\hfill\fi\ignorespaces}
\makeatother

\newcommand\Rey{\mbox{\textit{Re}}}  

\newsavebox{\astrutbox}
\sbox{\astrutbox}{\rule[-5pt]{0pt}{20pt}}

\newcommand\lap{ {\mathcal{L}}}
\newcommand\dz{ \frac{\partial}{\partial z}}

\newcommand\ls{}

\begin{document}

\title{Nonmodal Tollmien-Schlichting waves}

\author
{Joris\ls C.\ls G.\ls Verschaeve\thanks{Email address for correspondence: joris.verschaeve@gmail.com} \\
Noen vei 98, 0765 Oslo, Norway
}

\maketitle

\begin{abstract}
  The instability of flows via two-dimensional perturbations
  is analyzed theoretically and numerically
  in a nonmodal framework. The analysis is based on results obtained
  in [Verschaeve {\it et al.} (2018)] showing the inviscid character
  of the growth mechanism of these waves. In particular, it is shown that the
  formulation of this growth mechanism
  naturally reduces to the eigenvalue problem for the
  energy bound formulated
  by [Davis and von Kerczek (1973)]. The eigenvalue equation
  by [Davis and von Kerczek (1973)] thus allows for a broader interpretation.
  It provides the discrete growth rates for the
  base flow in question. In addition to this eigenvalue problem,
  a corresponding eigenvalue problem for the phase speed
  of the perturbations can be extracted from the equations found
  in [Verschaeve {\it et al.} (2018)]. 
  These two eigenvalue equations relate to the Hermitian and skew-Hermitian part, 
  respectively, of the nonmodal equations, cf.
  [Schmid (2007)]. In contrast to traditional Orr-Sommerfeld modal analysis, 
  the above eigenvalue equations define an orthogonal set of eigenfunctions allowing to
  decompose the perturbation into base perturbations with discrete growth rates and 
  frequencies. As a result of this decomposition, it can be shown that the evolution of 
  two-dimensional perturbations is governed by two mechanisms: A first one, 
  responsible for extracting 
  and returning energy from and to the base flow, in addition to viscous dissipation
  and, a second one, responsible for
  dispersing energy among the different base perturbations constituting the perturbation.
  As a general result, we show that 
  the stability of a flow is not only determined by the growth rates of the
  base perturbations, but it is also closely related to its ability
  to disperse energy away from the base perturbations with positive growth rates to the ones
  with negative growth rates. We illustrate the above results by means of
  three well known shear flows, Couette flow, Poiseuille flow,
  and the boundary layer flow under a solitary wave.
\end{abstract}

\section{Introduction}

Focus on two or three dimensional perturbations has shifted throughout the history of
research on hydrodynamic stability. Whereas, as a consequence of Squire's theorem,
modal stability analysis is focused on two-dimensional perturbations, nonmodal stability theory \cite{ButlerFarrell1992,TrefethenTrefethenReddyDriscoll1993,Schmid2007} has lead to an increased interest in three dimensional perturbations, in particular, the so called streamwise streaks, as they appear to have dominant amplifications during primary instability in canonical flows such as Couette, Poiseuille or Blasius. 

However, there are other shear flows, such as Stokes' second problem or the boundary layer flow under a solitary wave, which display superior amplifications of two-dimensional perturbations, as well in experiments as in modal and nonmodal analysis. These two-dimensional perturbations are of wave type, similar to the eigenfunctions of the celebrated Orr-Sommerfeld equation which are called Tollmien-Schlichting waves. Nevertheless, as stated in \cite{TrefethenTrefethenReddyDriscoll1993}, a stability analysis based on eigenvalue analysis, not taking into account the non-orthogonal nature of the system might miss the essential instability mechanism. \cite{ButlerFarrell1992} argue that instead of finding the most dangerous eigenvalues, 
it is more fruitful to search for the physical mechanisms leading to instability. 

The aim of the present discussion is precisely to investigate the physical instability mechanism of two-dimensional optimal perturbations, which can be thought of as a particular realization of a nonmodal Tollmien-Schlichting wave. In particular, we shall focus on the conditions which lead to decay, and weak or strong growth of nonmodal Tollmien-Schlichting waves.

The present investigation is a continuation of the nonmodal stability investigation
in \cite{VerschaevePedersenTropea2018}. \cite{VerschaevePedersenTropea2018} showed that the growth mechanism of two-dimensional perturbations
is inviscid, contrary to the common perception of a viscous growth mechanism.
In the present treatise, we shall, starting from the theoretical results obtained in \cite{VerschaevePedersenTropea2018}, derive an equation for the maximum growth rate attainable by any two-dimensional perturbation. This formula leads naturally to an eigenvalue problem equivalent to the eigenvalue problem in \cite{DavisKerczek1973} which has been derived by an alternative path via an energy bound on the Navier-Stokes equations.
This eigenvalue problem corresponds to a
higher order generalization of a Sturm-Louville eigenvalue problem. It defines an orthogonal/unitary set of eigenfunctions where the eigenvalues are the growth rates of the eigenfunctions. Next to this eigenvalue problem, an additional eigenvalue problem can be found whose eigenfunctions correspond to neutrally stable waves with a frequency given by the eigenvalues of this equation. As a consequence these two eigenvalue problems relate to the Hermitian and skew-Hermitian part, respectively, of the nonmodal equations. The eigenfunctions can be seen as base perturbations of the nonmodal Tollmien-Schlichting wave. These Tollmien-Schlichting base perturbations
allow to write the nonmodal equations in a Heisenberg form by means of infinite matrices. In this setting, the Hermitian part can be interpreted as a base Hamiltonian, whereas the skew-Hermitian part stands for the action of an additional potential distributing the energy onto different base states. Such systems are frequently found in quantum electrodynamics.

Following the above theoretical analysis, we perform numerical experiments for Couette flow, Poiseuille flow, and the boundary layer flow under a solitary wave. Although it is well known that streamwise streaks display amplifications several magnitudes larger than two-dimensional perturbations for the first two flow examples, it is nevertheless instructive to elucidate the above mechanisms for these base flows. The optimal nonmodal Tollmien-Schlichting wave results from an optimization of the energy transfer between stable and unstable Tollmien-Schlichting base perturbations, such that for a given point in time, the amplification at this point in time is largest. On the other hand, when the boundary layer flow under a solitary wave displays an adverse pressure gradient, the cascade of energy between Tollmien-Schlichting base perturbations is of less importance. Instead, the unstable Tollmien-Schlichting base perturbations experience exponential growth by extracting energy from the base flow with small transfer of energy to other base perturbations.

The present discussion is organized as follows. In section \ref{sec:problem}, we present the basic results and equations of hydrodynamic stability theory necessary for the present discussion. The results are presented in section \ref{sec:results} which is divided into two parts, one discussing the theoretical results and a second one presenting the numerical analysis. In section \ref{sec:conclusion}, we conclude the present discussion.

\section{Description of the problem} \label{sec:problem}

This section presents the basic equations on which the present analysis in section \ref{sec:results} is founded.\\

In the present treatise we shall consider steady and unsteady base flows $ \mathbf{U}_\base $ in
horizontal direction, with the wall normal direction in $ z $:
\be
\mathbf{U}_\base = U(z,t) \mathbf{e}_x . \label{eq:base}
\ee
We introduce a perturbation velocity $ \mathbf{u}' = (u',v',w' )$ in the 
streamwise, spanwise and wall normal direction, defined by:
\be
\mathbf{u}' = \left(u',v',w'\right) = \left(u_{ns},v_{ns},w_{ns} \right) 
- \left(U \left(z,t \right),0,0\right),
\ee
where $ (u_{ns},v_{ns},w_{ns} ) $ satisfies the Navier-Stokes equations. The
energy of the perturbation is given by:
\be
E_p = \frac{1}{2} \int \limits_V u'^2 + v'^2 + w'^2 \, dV,
\ee
which is integrated over the entire volume of interest $ V $. For time
dependent flows, \cite{DavisKerczek1973} derived
a bound for the perturbation energy for the Navier-Stokes equations:
\be
\frac{E_p(t)}{E_p(t_0)} \le \exp \int \limits_{t_0}^t
\mu(t') \, dt',
\ee
where $ \mu $ is the largest eigenvalue of the following linear system:
\bea
\frac{1}{\Rey} \Delta \mathbf{u}' - \mathbf{S}_\base(t) \cdot
\mathbf{u}' - \nabla p &=& \frac{1}{2} \mu \mathbf{u}' \label{eq:davis1} \\
\nabla \cdot \mathbf{u}' &=& 0, \label{eq:davis2}
\eea
where the tensor $ \mathbf{S}_\base $ is the rate of strain tensor
given by the base flow, equation (\ref{eq:base}). Compared to the equations in 
\cite{DavisKerczek1973}, a missing factor of $ 1/2 $ has been accounted for in (\ref{eq:davis1}-\ref{eq:davis2}).
As the rate of strain tensor depends on time $ t $,
the eigenvalue $ \mu $ is also time dependent.
If $ \mu < 0 $ for all times, then the flow is
monotonically stable for this Reynolds number, meaning that
all perturbations always decay.
As the base flow is independent of $ x $ and $ y $, we consider a single Fourier
component of $ \mathbf{u}' $:
\be
(u',v',w')(x,y,z,t) = (u,v,w)(z,t) \exp {\rm i} \left( \alpha x + \beta y \right).
\ee
This allows us to eliminate $ p $ from the equations (\ref{eq:davis1}-\ref{eq:davis2}), resulting into 
\bea
\frac{1}{\Rey} \lap^2 w + \frac{ {\rm i } \alpha}{2}
\left\{ D^2 U w + 2 D U D w \right\} + \frac{{\rm i} \beta}{2}  D U \zeta  &=& \frac{1}{2}\mu \lap w,
\label{eq:davis1a}\\
-\frac{1}{\Rey} \lap \zeta - \frac{ {\rm i} \beta}{2} D U w &=& \frac{1}{2}\mu (-\zeta) \label{eq:davis2a}
\eea
where $ \lap $ is the Laplacian defined by:
\be
\lap = D^2 -k^2 .
\ee
We introduced the following shorthand notations:
\bea
D &=& \frac{\partial}{\partial z}, \\
k^2 &=& \alpha^2 + \beta^2.
\eea
The system of four equations (\ref{eq:davis1}-\ref{eq:davis2}),
has been reduced to two, by means of the normal vorticity component $ \zeta $:
\be
\zeta 
= {\rm i} \left( \alpha v - \beta u \right).
\ee
As mentioned above, the system (\ref{eq:davis1})-(\ref{eq:davis2}) and the system (\ref{eq:davis1a})-(\ref{eq:davis2a}) result from a bound on the energy of the perturbation starting from the Navier-Stokes equations. As we shall see in section \ref{sec:results}, an equivalent eigenvalue problem can be obtained by introducing a parabolized stability equation ansatz for the perturbation and searching for the maximum growth rate.\\

Nonmodal stability analysis is based on the linearized Navier-Stokes
equations, which can be written in the present setting as follows,
\bea
\left( \frac{\partial}{\partial t} + {\rm i} \alpha U - \frac{1}{\Rey} \mathcal{L}
\right) \mathcal{L} w
- {\rm i} \alpha w D^2 U  &=& 0, \label{eq:sys1} \\
\left( \frac{\partial}{\partial t} + {\rm i} \alpha U  - \frac{1}{\Rey} \mathcal{L}
\right) \zeta - {\rm i} \beta w D U 
&=& 0. \label{eq:sys2}
\eea
We refer to \cite{SchmidHenningson2001,Schmid2007} for a thorough derivation
of equations (\ref{eq:sys1}) and (\ref{eq:sys2}). Given an initial
perturbation $ (w_0,\zeta_0 ) $ at time $ t_0 $, equations 
(\ref{eq:sys1}) and (\ref{eq:sys2}) can be integrated to
obtain the temporal evolution of $ (w,\zeta)$ for $ t > t_0 $.
Nonmodal theory formulates the stability problem as finding
the initial condition $(w_0,\zeta_0 ) $ maximizing the perturbation
energy $ E_p(t) $ of $ (w,\zeta) $ at time $ t > t_0 $. 
This perturbation energy $ E_p $ is the sum
of two contributions, one from the wall normal component $ w $ 
and one from the normal vorticity component $ \zeta$:
\be
E_p(t) = E_w(t) + E_\zeta(t) = 
\frac{1}{2} \int \limits_a^b \frac{1}{k^2} \left| D w \right|^2 + \left| w \right|^2 \, dz 
+ \frac{1}{2} \int \limits_a^b \frac{1}{k^2} \left| \zeta \right|^2 \, dz,
\label{eq:energyNonmodal}
\ee
where $ a = -1 $ and $ b = 1$ for the enclosed base flows and $ a = 0 $ and $ b = \infty $ for the boundary layer flow in the present discussion.

The optimization problem can then be formulated by maximizing $ E_p$ for
a perturbation $ (w,\zeta) $ satisfying (\ref{eq:sys1}) and (\ref{eq:sys2})
and having an initial energy $ E_p(t_0) $. One way of solving this
optimization problem is 
by means of the adjoint equation as in \cite{LuchiniBottaro2014}. Another
approach for finding the optimal perturbation, which is employed in
the present treatise, consists in formulating
the discrete problem first and computing the fundamental solution matrix $ \mathbf{X}(t,t_0) $ of the
system of ODEs, cf. references \cite{TrefethenTrefethenReddyDriscoll1993,SchmidHenningson2001,Schmid2007} for details. The energy $ E_p $ is then
given in terms of $ \mathbf{X} $ and the initial condition. Details of
the implementation are given in \cite[appendix A]{VerschaevePedersenTropea2018}. 
By computing $ E_p(t) $ one way or the other, we can compute
the amplification $ G $ from time $ t_0 $ to $ t $ 
of the optimal perturbation for wave numbers $ \alpha $ and $ \beta$:
\be
G(\alpha,\beta,t_0,t,\Rey) = \max_{ (w_0,\zeta_0) } \frac{ E_p(t) }{E_p(t_0) }. \label{eq:amplification}
\ee
We remark that the initial condition $ (w_0,\zeta_0 ) $
from which the optimal perturbation starts,
might be different for each point in time $ t $,
when tracing $ G $ as a function of $ t $, cf. section \ref{sec:results}.
The maximum amplification $ G_{\max}(\Rey)$, which can be reached
for a given Reynolds number $ \Rey $, is obtained by
maximizing $ G$ over time, initial time and wavenumbers:
\be
G_{\max} = \max_{\alpha,\beta,t_0,t} G. \label{eq:maximumA}
\ee
In the following, we shall distinguish between the following three types of perturbations:
\vspace{2mm}
\begin{itemize}
\item Streamwise streaks.\\
These are perturbations independent of the streamwise coordinate $ x $. They
can be computed by setting $ \alpha = 0 $. For this case, the normal velocity component $ w $ is a slowly decaying function entering equation (\ref{eq:sys2}) as a source term together with $ U $.
\item Two dimensional perturbations.\\
If $ \beta = 0 $, the governing equations (\ref{eq:sys1}) and (\ref{eq:sys2})
are decoupled. All possible growth is restricted to $ w $, as the normal vorticity $ \zeta $ only displays decay.
\item Oblique perturbations.\\
These are all perturbations with $ \alpha \neq 0 $ and $ \beta \neq 0 $. 
\end{itemize}
\vspace*{1ex}
The present treatise focuses on perturbations of the second type which we shall call nonmodal Tollmien-Schlichting waves due to their wave-type character. The third perturbation type can be considered superpositions of nonmodal Tollmien-Schlichting waves and streamwise streaks. For this type of perturbation, the nonmodal Tollmien-Schlichting wave travels in the plane spanned by the wave number vector $ \mathbf{k} = (\alpha,\beta) $ and $ \mathbf{e}_z $.

For streamwise streaks ($\alpha=0$), on the other hand, a scaling argument as in \cite{Gustavsson1991,SchmidHenningson2001} allows us to rewrite equations (\ref{eq:sys1})
and (\ref{eq:sys2}):
\bea
\left( \frac{\partial}{\partial \tau}  - \mathcal{L}
\right) \mathcal{L} {w}
 &=& 0, \label{eq:sys1streak2} \\
\left( \frac{\partial}{\partial \tau} - \mathcal{L}
\right) \tilde{\zeta} - {\rm i} \beta {w} \dz U 
&=& 0, \label{eq:sys2streak2}
\eea
where $ \tau = t/\Rey $ is a slowly varying time scale. The normal vorticity
$ \tilde{\zeta} $ is scaled by $ \Rey $:
\be
\tilde{\zeta} = \frac{1}{\Rey} \zeta(z,\tau) .
\ee
Equation (\ref{eq:sys1streak2}) corresponds to slow
viscous damping of $ w $, which also holds for the homogeneous part of
equation (\ref{eq:sys2streak2}) for $ \tilde{\zeta} $. The second term in (\ref{eq:sys2streak2}) represents
a forcing term which varies on the temporal scale of
the base flow. Therefore, streamwise streaks display
temporal variations on the time scale of the base flow.\\

A similar scaling argument, however, does not hold for nonmodal Tollmien-Schlichting waves ($ \alpha > 0 $),
as they often oscillate on smaller (faster) time scales than the base flow.\\

The theoretical analysis of nonmodal Tollmien-Schlichting waves is greatly facilitated by employing a parabolized stability equation ansatz. As such the parabolized stability equation has been derived as a numerical method in \cite{BertolottiHerbertSpalart1992}. However, in \cite{VerschaevePedersenTropea2018} it has been adapted for a theoretical analysis of the nonmodal stability of the boundary layer under a solitary wave.

In this approach, the normal velocity component $ w $ is decomposed into a shape function $ \tilde{w} $ and an exponential factor: 
\be
w = \tilde{w}(z,t) \exp -{\rm i} \int \limits_{t_0}^t \Omega(t') \, dt', \label{eq:pseDecomposition}
\ee
where the real part of $ \Omega $ accounts for
the oscillatory character of $ w $ and the imaginary part
of $ \Omega $ is the growth rate of the perturbation. We remark that in \cite{VerschaevePedersenTropea2018}, the factor $ - {\rm i} $ has been absorbed into the definition of $ \Omega $.\\

In order to define the shape function $ \tilde{w} $ univocally, all growth is restricted to $ \Omega $. Somewhat different to \cite{BertolottiHerbertSpalart1992},
a normalization condition is defined on the entire kinetic energy $ \tilde{E}_w $ of the shape function $ \tilde{w} $ :
\be
\tilde{E}_w = \frac{1}{2} \int \limits_a^b \frac{1}{k^2} | D\tilde{w} |^2 + | \tilde{w} |^2 \, dz.
\ee
The normalization constraint can thus be written as:
\be
 \int \limits_a^b  \tilde{w}^\dagger \mathcal{L}  \frac{\partial \tilde{w} }{\partial t} \, dz = 0, \label{eq:constraint}
 \ee
 from which it follows that we can require that $ \tilde{E}_w $ is unity for all times
 \be
\tilde{E}_w = 
-\frac{1}{2k^2}\int \limits_a^b \tilde{w}^\dagger \mathcal{L} \tilde{w} \, dz =  1.
\label{eq:constraintEnergy}
\ee
Equation (\ref{eq:sys1}) becomes then:
\be
\partial_t \mathcal{L} \tilde{w} - {\rm i} \Omega \mathcal{L} \tilde{w}
= \frac{1}{\Rey } \mathcal{L}^2 \tilde{w} + {\rm i} \alpha 
\left( D^2 U - U \mathcal{L} \right) \tilde{w} \label{eq:pseGoverning}
\ee
Multiplying by $ \tilde{w}^\dagger $ and integrating in $ z $, leads to
a formula for $ \Omega $:
\be
\Omega = -\frac{\rm i }{2 k^2 \Rey} \int \limits_a^b \tilde{w}^\dagger \mathcal{L}^2 \tilde{w} \, dz
+ \frac{ \alpha }{2 k^2} 
\int \limits_a^b \tilde{w}^\dagger D^2 U \tilde{w} - \tilde{w}^\dagger U \mathcal{L} \tilde{w} \, dz \label{eq:omegaPSE}
\ee
In order to facilitate the notation, we shall write for the real and imaginary parts of $ \Omega $:
\be
\Omega = \omega + {\rm i} \sigma.
\ee
From equation (\ref{eq:omegaPSE}), \cite{VerschaevePedersenTropea2018} derived a formula for the growth rate $ \sigma $:
\bea
\sigma &=&  - \frac{1}{2 k^2 \Rey}
\int \limits_a^b | \mathcal{L} \tilde{w}|^2
\, dz  -  \frac{{\rm i} \alpha}{4 k^2}
\int \limits_a^b D U \left\{ \tilde{w}^\dagger D \tilde{w} - D \tilde{w}^\dagger \tilde{w}  \right\} \, dz \\
&=& 
 - \frac{1}{2 k^2 \Rey}
\int \limits_a^b | \mathcal{L} \tilde{w}|^2
\, dz  +  \frac{\alpha}{2 k^2}
\int \limits_a^b D U \left\{ \tilde{w}_r D \tilde{w}_i - \tilde{w}_i D \tilde{w}_r \right\} \, dz \label{eq:growth}
\eea
As \cite{VerschaevePedersenTropea2018} noted, the first term on the right hand side represents viscous dissipation and is always negative. The
second term, however, can, depending on $ U $ and $ \tilde{w} $, be positive or negative.
Only when this term is positive and in magnitude larger than the viscous dissipation,
growth of $ E_w $ can be observed. As is immediately evident, the second term is multiplied by $ \alpha = \sqrt{ k^2 - \beta^2 } $, which is maximum for $ \beta = 0 $ for a given $ k $. Therefore, nonmodal Tollmien-Schlichting waves will display larger growth rates when aligned with the base flow for cases where the energy contribution by the normal vorticity component is negligible. This result for nonmodal Tollmien-Schlichting waves corresponds to Squire's theorem for modal perturbations. We remark that for many flow situations, the normal vorticity $ \zeta $ does not vanish and that larger energy growth can often be obtained for $ E_\zeta $ than for $ E_w $. Under these circumstances the superposition (or the pure streamwise streaks) will have a larger amplification than the nonmodal Tollmien-Schlichting wave aligned with the flow.\\

Performing some algebraic manipulations, \cite{VerschaevePedersenTropea2018} expressed $ \sigma $ as a function of the rate of strain $ \mathbf{S}_\base $ for two-dimensional perturbations ($ \beta = 0$):
\bea
\sigma & =& - \frac{1}{2 \alpha^2 \Rey}
\int \limits_a^b | \mathcal{L} \tilde{w}|^2
\, dz  - \frac{ 1 }{2}
\int \limits_a^b  \left( \tilde{u}^\dagger , \tilde{w}^\dagger \right)
\mathbf{S}_\base
 \left( \begin{array}{c} \tilde{u} \\ \tilde{w} \end{array} \right), \label{eq:tilting}
 \eea
 where $ \tilde{u} $ is the horizontal velocity component given by:
 \be
{\rm i} \alpha \tilde{u} = - D \tilde{w}. 
 \ee
 In the oblique case, a similar relation holds for the projection of $ \mathbf{S}_\base $ onto
 the wave number vector $ \mathbf{k} = ( \alpha , \beta ) $, cf. \cite{VerschaevePedersenTropea2018}. Formula (\ref{eq:tilting}) corresponds to a tilt of the velocity vector $ \left( \tilde{u} , \tilde{w} \right)^T $ by the
 rate of strain tensor $ \mathbf{S}_\base $ which can be seen as the
 Orr-mechanism in a nonmodal framework. As
 \cite{VerschaevePedersenTropea2018} concluded,
 the growth mechanism itself is always inviscid, holding also
 for modal Tollmien-Schlichting waves which
 are commonly thought of as slow viscous instabilities, cf. for example
 \cite{Jimenez2013} and \cite{BrandtSchlatterHenningson2004}.
 Whether growth of two-dimensional perturbations is fast or slow is, as
 formula (\ref{eq:tilting}) suggests, primarily
 a property of the base flow profile $ U $.\\

In the following, we shall further develop the formalism developed in \cite{VerschaevePedersenTropea2018} and show how it allows us to deepen the theoretical understanding of nonmodal Tollmien-Schlichting waves.\\

Concerning the generation of numerical results, we employed the same numerical framework as in \cite{VerschaevePedersenTropea2018}. For the solution of the eigenvalue equations, a discretization based on Shen-Legendre polynomials is employed \cite{Shen1994}. The time dependent equations are discretized by means of Shen-Chebyshev polynomials \cite{Shen1995} and integrated via a Runge-Kutta integrator, see \cite{VerschaevePedersenTropea2018} for implementation details and for verification and validation.

 \section{Results} \label{sec:results}

 We shall first present theoretical derivations in section \ref{sec:theory} before going over to numerical results in section \ref{sec:numerics}.

 \subsection{Theoretical considerations} \label{sec:theory}

In the following, we shall often use the energy scalar product for two functions $ \phi $ and $ \psi $ satisfying the boundary conditions at $ a $ and $ b $:
\be
\left\langle \phi , \psi \right\rangle = \frac{1}{2k^2} \int \limits_a^b D \phi^\dagger D \psi + k^2 \phi^\dagger \psi \, dz.
\ee
 Based on the definition of $ \tilde{w} $, equation (\ref{eq:pseDecomposition}), the energy $ E_w $ of the perturbation can be expressed by means of the growth rate $ \sigma $:
 \be
 E_w = \frac{1}{2} \int \limits_a^b \frac{1}{k^2} | D\tilde{w} |^2 + | \tilde{w} |^2 \, dz \exp 2 \int \limits_{t_0}^t \sigma(t') \, dt' =
 \exp 2 \int \limits_{t_0}^t \sigma(t') \, dt'. \label{eq:energyPSE}
 \ee
 This allows us to find a bound on $ E_w $ by searching for the shape function
 $ \tilde{w} $ which maximizes $ \sigma $ for any point in time $ t $, ie. we have the following variational problem:
 \be
\max_{\tilde{w}} \sigma = \max_{\tilde{w}} - \frac{1}{2 k^2 \Rey}
\int \limits_a^b \mathcal{L} \tilde{w}^\dagger \mathcal{L} \tilde{w}
\, dz  +   \frac{\alpha}{2 k^2}
\int \limits_a^b D U \left\{ \tilde{w}_r D \tilde{w}_i - \tilde{w}_i D \tilde{w}_r \right\} \, dz.  \label{eq:maximumGrowthRate} 
\ee
However, $ \tilde{w} $ has to satisfy the constraint that its energy is unity, which leads to the following Lagrangian:
\bea
{L_\gamma[\tilde{w}_r,\tilde{w}_i,\lambda]} &=&
- \frac{1}{2k^2 \Rey }
\int \limits_a^b \mathcal{L} \tilde{w}^\dagger \mathcal{L} \tilde{w}
\, dz  +  \frac{\alpha}{2 k^2}
\int \limits_a^b D U \left\{ \tilde{w}_r D \tilde{w}_i - \tilde{w}_i D \tilde{w}_r \right\} \, dz  \nonumber \\
& & \quad + \lambda \left( 1 - \frac{1}{2} \int \limits_a^b \frac{1}{k^2} | D\tilde{w} |^2 + | \tilde{w} |^2 \, dz \right) \label{eq:Lagrangian1}
\eea
Stationarity with respect to $ \tilde{w} $ leads to the following 
eigenvalue equation:
\be
\frac{1}{\Rey} \mathcal{L}^2 \tilde{w}
+ \frac{ {\rm i} \alpha}{ 2 } \left( 2 DU D \tilde{w} + D^2 U \tilde{w} \right)
 =  \lambda \mathcal{L} \tilde{w}, \label{eq:davisPSE} 
\ee
where the eigenvalue $ \lambda $ corresponds to $ \sigma $, which can be seen by multiplying equation (\ref{eq:davisPSE}) by $ \tilde{w}^\dagger $ and integrating in wall normal direction. It is straightforward to verify that the differential operator on the left hand side of (\ref{eq:davisPSE}) is a Hermitian operator. The eigenvalues are thus real and the eigenfunctions are orthogonal with respect to the energy scalar product. Seen the definition of the energy (\ref{eq:energyPSE}), equation (\ref{eq:davisPSE}) is equivalent to the eigenvalue system (\ref{eq:davis1a}-\ref{eq:davis2a}), in the case $ \beta = 0 $, previously found by \cite{DavisKerczek1973}. Formulating a bound on the energy of the perturbation or alternatively formulating a bound on its growth rate via the present parabolized stability equation formulation produces equivalent systems, which supports the approach developed in \cite{VerschaevePedersenTropea2018}. \\

For $ \Rey \rightarrow \infty $, the variational problem (\ref{eq:maximumGrowthRate})
reduces to
\be
\max_{\tilde{w}} \sigma = \max_{\tilde{w}}  \frac{\alpha}{2 k^2}
\int \limits_a^b D U \left\{ \tilde{w}_r D \tilde{w}_i - \tilde{w}_i D \tilde{w}_r \right\} \, dz, 
\ee
which, when using Euler's formula:
\be
\tilde{w}(z) = r(z) \exp \theta(z), \label{eq:euler}
\ee
becomes
\be
\max_{r,\theta} \sigma = \frac{\alpha}{2 k^2} \int \limits_a^b r^2 D\theta D U   \, dz. 
\ee
At extremum, while respecting constraint (\ref{eq:constraintEnergy}), we obtain the following eigenvalue problem:
\bea
D \theta &=& \frac{\alpha}{2\lambda} D U \label{eq:inviscidDavis1} \\
\lambda^2 \left( D^2 - k^2 \right) r + \frac{\alpha^2}{4} \left( D U \right)^2 r &=& 0 . \label{eq:inviscidDavis2}
\eea
Using equation (\ref{eq:inviscidDavis1}), the maximum for the growth rate 
can thus be written as:
\be
\max_{r,\theta} \sigma = \max_{r} \sigma = 
\max_{r} \frac{\alpha}{2 k} \sqrt{ \int \limits_a^b r^2 \left( D U\right)^2   \, dz }. \label{eq:maximumGrowthRateInviscid} 
\ee
As can be seen from equation (\ref{eq:inviscidDavis2}), in the inviscid case, the eigenvalues $ \lambda $ come in pairs of positive and negative values, corresponding to growth and decay respectively.
More interestingly, however, is the fact, that the phase change $ D \theta $ of the eigenfunctions of system (\ref{eq:inviscidDavis1}-\ref{eq:inviscidDavis2}) is proportional to the rate of strain of the base flow, cf. equation (\ref{eq:inviscidDavis1}). This can be interpreted as a {\em resonance} mechanism of the system selecting the perturbation whose phase change matches the strain rate best. \\

However, as we shall see in the following, equation (\ref{eq:davisPSE}) and equations (\ref{eq:inviscidDavis1}) and (\ref{eq:inviscidDavis2}) only describe one aspect of the physical mechanism of growth of nonmodal Tollmien-Schlichting waves. In order to obtain a complete picture of the growth mechanism, we return to the constraint, equation (\ref{eq:constraint}), on which the present parabolized stability equation formalism is founded. Writing out the real and imaginary part of constraint (\ref{eq:constraint}), we find:
\bea
\lefteqn{\frac{1}{2} \frac{d}{dt} \int \limits_a^b
  (D\tw_r)^2 + (D\tw_i)^2 + k^2 (\tw_r^2 + \tw_i^2) \, dz} \nonumber \\
& & + {\rm i} \int \limits_a^b \left( - \tw_i \mathcal{L} \frac{\partial}{\partial t} \tw_r
+ \tw_r \mathcal{L} \frac{\partial}{\partial t} \tw_i \right) \, dz = 0.\label{eq:constraintRealAndImag}
\eea
The real part of equation (\ref{eq:constraintRealAndImag}) corresponds to the conservation of energy imposed on $ \tw $ and which ultimately defines the growth rate $ \sigma $, equation (\ref{eq:growth}). On the other hand, using equation (\ref{eq:pseGoverning}), the constraint on the imaginary part can be reformulated as:
\bea
\omega = \frac{\alpha}{2k^2} \int \limits_a^b \frac{1}{2} (D^2 U) |\tw|^2
+ U \left( | D \tw |^2 + k^2 | \tw |^2 \right) \, dz, \label{eq:dispersion}
\eea
which coincides with the imaginary part of equation (\ref{eq:omegaPSE}). This implies that the present parabolized stability equation formalism could have been derived by only formulating a constraint on the real part of equation (\ref{eq:constraintRealAndImag}). 
In other words, imposing the constraint (\ref{eq:constraintRealAndImag}) on the imaginary part is redundant. Equation (\ref{eq:dispersion}) is a formula for the frequency of nonmodal Tollmien-Schlichting waves. If the dependence of $ \tilde{w} $ on $ \alpha $ had been known, equation (\ref{eq:dispersion}) would give us an explicit dispersion relation for the frequency $ \omega $ as a function of the wavenumber $ \alpha $. \\
Given equation (\ref{eq:dispersion}), the phase speed $ c $ of the nonmodal Tollmien-Schlichting wave can be written as:
\be
c = \frac{\omega}{\alpha} = \frac{1}{2k^2} \int \limits_a^b \frac{1}{2} (D^2 U) |\tw|^2
+ U \left( | D \tw |^2 + k^2 | \tw |^2 \right) \, dz. \label{eq:phaseSpeed}
\ee
It consists of two terms. The first one can be seen as a weighted integral of $ D^2 U $, with $ | \tw |^2/4k^2 $ being the weight function. The second term is a weighted integral of $ U $ with the energy density of $ \tw $ as a weight. As the energy density of $ \tw $ sums to unity, the weighted integral of $ U $ gives in fact a mean velocity. This term adds more weight to the value of $ U $ where the energy density of the perturbation is largest. As such the second term indicates that the nonmodal Tollmien-Schlichting wave travels with a wave speed comparable to the base flow velocity, if the influence of the first term is negligible. 

Equation (\ref{eq:dispersion}) can also be interpreted as the solution to the constrained optimization problem defined by the following Lagrangian:
\bea
L[\tw_r, \tw_i , \nu ] &=&  \frac{\alpha}{2k^2} \int \limits_a^b \frac{1}{2} (D^2 U) |\tw|^2
+ U  \left( | D \tw |^2 + k^2 | \tw |^2 \right) \, dz \nonumber \\
& & \quad + \nu \left( 1 - \frac{1}{2} \int \limits_a^b \frac{1}{k^2} | D\tilde{w} |^2 + | \tilde{w} |^2 \, dz \right). \label{eq:Lagrangian2}
\eea
We remark that $ \nu $ is not the kinematic viscosity, but a Lagrangian multiplier in order to satisfy constraint (\ref{eq:constraintEnergy}).  
At stationarity, we obtain the following eigenvalue equation:
\be
{\alpha} \left( \frac{1}{2} D^2 U \tw - D U D \tw - U \mathcal{L} \tw  \right) = - \nu \mathcal{L} \tw.  \label{eq:eigenFrequency}
\ee
Multiplying this equation by $ \tw^\dagger $ and integrating in $ z $, we find that at extremum $ \nu $ corresponds to $ \omega $. A straightforward calculation allows to verify that equation (\ref{eq:eigenFrequency}) is self-adjoint and therefore its eigenvalues are real and its eigenfunctions orthogonal with respect to the energy scalar product. \\

Turning again to Euler's formula, equation (\ref{eq:euler}), we obtain for the eigenvalue problem (\ref{eq:eigenFrequency}):
\bea
D \theta &=& 0 \label{eq:frequencyEuler1} \\
\frac{\alpha}{2} \left( D^2 U r - 2 D \left( U D r \right) + 2 k^2 U r \right) &=& \nu \left( - D^2 r + k^2 r \right)
\eea
Equation (\ref{eq:frequencyEuler1}) corresponds to the fact, that the system is self-adjoint (in the sense that it can be written as a real and not a complex equation), opposed to the above truly Hermitian one, equation (\ref{eq:davisPSE}). Therefore, 
contrary to the eigenfunctions of equation (\ref{eq:davisPSE}), the eigenfunctions of (\ref{eq:eigenFrequency}) can be normalized to be real, ie. $ \theta = 0$. Equation (\ref{eq:frequencyEuler1}) also contrasts with the formula for the phase of the eigenfunctions with maximum growth rate in the inviscid case, equation (\ref{eq:inviscidDavis1}), indicating that in general the optimum of one system cannot be the optimum of the other. \\

Eigenvalue problem (\ref{eq:davisPSE}) is related to the buckling problem of a thin rod \cite{VentselKrauthammer2001}. The eigenvalues $ \lambda_i $, equation (\ref{eq:davisPSE}), appear as a decreasing sequence:
\be
\lambda_0 > \lambda_1 > \ldots > \lambda_i > \ldots. \label{eq:eigenvalueLambda}
\ee
As mentioned in \cite{DavisKerczek1973}, growth of nonmodal Tollmien-Schlichting waves is possible only if at least $ \lambda_0 $ is larger than zero. As can be seen from equation (\ref{eq:maximumGrowthRate}), the viscous contribution, ie. the first term on the left hand side of (\ref{eq:maximumGrowthRate}), cannot be bounded from below, leading to discrete eigenvalues towards $ -\infty $, at least for bounded domains. An upper bound $ \lambda_{\max} > 0 $ for $ \lambda $ is given in \cite[formula (2.8)]{DavisKerczek1973} for bounded domains. However, even for unbounded domains, we can use equation (\ref{eq:maximumGrowthRateInviscid}) to find an upper bound for the second term in (\ref{eq:maximumGrowthRate}), as long as $ \int_a^b ( D U )^2 dz < \infty $, which is the case for all flows considered in the present discussion. Using H\"olders inequality, we find:
\bea
\lambda^2 & \leq & \frac{\alpha^2}{4 k^2} \int \limits_a^b r^2 \left( D U\right)^2   \, dz \\
 &\leq& \frac{\alpha^2}{4 k^2} \int \limits_a^b r^2 \, dz 
\int \limits_a^b \left( D U\right)^2 \, dz \\
& \leq & \underbrace{\frac{\alpha^2}{2 k^2}  \int \limits_a^b \left( D U\right)^2 \, dz}_{ =: \lambda_{\max}^2}, \label{eq:lambdaMax}
\eea
where we have used the fact that the energy of the perturbation is unity and we thus have the following bound for the integral of $ r^2 $:
\be
\frac{1}{2}  \int \limits_a^b r^2 \, dz 
= \frac{1}{2} \int \limits_a^b | \tw |^2 \, dz \leq \frac{1}{2k^2} \int \limits_a^b | D\tw |^2 + k^2 | \tw |^2 \, dz = 1.
\ee
In general, the mathematical theory on higher order Sturm-Liouville problems with a second order differential operator in the term multiplied by the eigenvalue is rather limited \cite{Chatterji1998}. A mathematical proof of property (\ref{eq:eigenvalueLambda}) of the discrete spectrum of equation (\ref{eq:davisPSE}) is still an open question. In addition, for unbounded domains, a mathematical analysis of its discrete and continuous spectrum is still missing. However, these questions surpass the scope of the present discussion. The numerical results in \ref{sec:numerics} give an indication that a discrete spectrum with property (\ref{eq:eigenvalueLambda}) exists under some conditions even for unbounded domains.

Equation (\ref{eq:eigenFrequency}) in this respect is even more exceptional in the sense that the term multiplied by the eigenvalue is as the left hand side, a second order differential operator. The eigenvalues of (\ref{eq:eigenFrequency}) are observed to lie in a specific range, cf. section \ref{sec:numerics}:
\be
\nu_i \in [ \omega_{\min} , \omega_{\max} ]. \label{eq:boundOnFrequency}
\ee
From a physical point of view, this is sensible, as it should not be possible for a nonmodal Tollmien-Schlichting to travel with infinite speed. Property (\ref{eq:boundOnFrequency}) can be proven the following way.\\

For the boundary layer flow in the present discussion, we have 
\be
\lim_{z \rightarrow \infty} U(z,t) = U_b(t),
\ee
where $ U_b $ is not necessarily zero. In this case the integral
$ \int_0^\infty |U|  dz $ is not defined. However, we assume that
when subtracting $ U_b $ from $ U $, its integral is bounded, ie. we have:
\be
\int \limits_0^\infty \left| U - U_b \right| \, dz < \infty. 
\ee
Using equation (\ref{eq:dispersion}), we can find a bound for $ | \omega | $:
\bea
\frac{2k^2}{\alpha} | \omega| & \leq & 
\left| \int \limits_a^b \frac{1}{2} (D^2 U) |\tw|^2
+ U \left( | D \tw |^2 + k^2 | \tw |^2 \right) \, dz \right| \\
& \leq & 
\left| \int \limits_a^b \frac{1}{2} (D^2 U) |\tw|^2 \, dz \right| 
+ \left| \int \limits_a^b U \left( | D \tw |^2 + k^2 | \tw |^2 \right) \, dz \right| \\
& = & 
\left| \int \limits_a^b \frac{1}{2} (D^2 U) |\tw|^2 \, dz \right| 
+ \left| \int \limits_a^b \left(U - U_b + U_b\right) \left( | D \tw |^2 + k^2 | \tw |^2 \right) \, dz \right| \\
& \leq &  \int \limits_a^b \left| D^2 U \right| \,dz
\int \limits_a^b \frac{1}{2} |\tw|^2 \, dz 
+ \int \limits_a^b \left| U - U_b\right| \,dz
\int \limits_a^b \left( | D \tw |^2 + k^2 | \tw |^2 \right) \, dz  \nonumber \\
&   & \quad \quad +| U_b | \int \limits_a^b \left( | D \tw |^2 + k^2 | \tw |^2 \right) \, dz  \\
& \leq & \int \limits_a^b \left| D^2 U \right| \,dz + 2 k^2 \int \limits_a^b \left| U - U_b\right| \,dz + 2k^2 | U_b |,
\eea
where we have made use of H\"older's inequality. A consequence of the boundedness of the eigenvalues of (\ref{eq:eigenFrequency}) is that we do not have a hierarchy of the eigenvalues as in (\ref{eq:eigenvalueLambda}). This manifests itself during the numerical calculations in the fact that when increasing the numerical resolution, the eigenvalues for finer resolutions appear in between the ones for coarser resolutions. \\

On the other hand, as the eigenvalues of equation (\ref{eq:davisPSE}) appear in a decreasing sequence, equation (\ref{eq:davisPSE}) allows us to define a proper set of orthonormal eigenfunctions with respect to the energy scalar product:
\be
\left\{\lambda_i, \phi_i \right\}, \quad i = 0, 1, 2, \ldots.
\ee
In the following, we shall call the functions $ \phi_i $ VKD-modes in honor of Von Kerczek and Davis who were the first to derive eigenvalue system (\ref{eq:davis1}-\ref{eq:davis2}). Prescinding from the continuous spectrum in the case of unbounded domains, the nonmodal Tollmien-Schlichting wave can then be expanded onto the VKD-modes:
\be
w = \sum \limits_{i=0} c_i \phi_i,
\ee
where
\be
c_i = \left\langle\phi_i , w \right\rangle.
\ee
In quantum mechanics, the VKD-modes would correspond to the base states of a quantum mechanical system with $ | c_i |^2 $ its probability to be found in this state. The analogy is not complete, as in the present case $ | c_i |^2 $ is the energy of the VKD-mode. However, the VKD-modes can be seen as the base perturbations of the governing system (\ref{eq:sys1}).
Defining $ \mathbf{c} = ( \ldots, c_i , \ldots )^t $ as the coefficient vector,
we can write the governing equation (\ref{eq:sys1}) in Heisenberg form:
\be
\frac{d}{dt}\mathbf{c} + \mathbf{F} \mathbf{c} = \mathbf{\Lambda} \mathbf{c} - {\rm i} \mathbf{N} \mathbf{c}. \label{eq:Heisenberg0}
\ee
The matrices in equation (\ref{eq:Heisenberg0}) have the following meaning. Matrix $ \mathbf{F} $ accounts for the temporal change of the VKD-modes. Its elements are defined by
\be
F_{ij} = \left\langle \phi_i , \frac{\partial}{\partial t} \phi_j \right\rangle. 
\ee
The base perturbations $ \phi_i $ result from an eigenvalue problem for each point in time. Therefore, they display temporal variations on the same scale as the base flow. In order to trace $ \phi_i $ in time, a mapping of the VKD-modes between different points in time needs to be established. This poses some challenges as shall be elaborated more in detail in appendix \ref{sec:modeContinuation}.
For steady base flows the matrix $ \mathbf{F} $ vanishes. Many boundary layer flows are governed by two well separated time scales. For the boundary layer flow under a solitary wave, the base flow varies on a slow time scale, namely $ \Rey/2 $, whereas the nonmodal Tollmien-Schlichting wave varies on a fast time scale. As $ \mathbf{F} $ varies on the same scale as the base flow, we can neglect it for higher Reynolds numbers. This is similar to neglecting nonparallel effects in boundary layers developing in streamwise direction. In order to ease the discussion, we shall only consider the cases where $ \mathbf{F} $ is negligible in the following. The governing equation becomes then
\be
\frac{d}{dt}\mathbf{c} = \mathbf{\Lambda} \mathbf{c} - {\rm i} \mathbf{N} \mathbf{c}. \label{eq:Heisenberg}
\ee
The matrix $ \mathbf{\Lambda} $ is a diagonal matrix with the growth rates on its diagonal:
\be
\Lambda_{ij} = \lambda_i \delta_{ij} .
\ee
The elements of matrix $ \mathbf{N} $ are defined as follows:
\be
N_{ij} = \frac{\alpha}{2k^2}
\int \limits_a^b \frac{1}{2} D^2 U \phi_i^\dagger \phi_j
+ U \left( D\phi_i^\dagger D \phi_j + k^2 \phi_i^\dagger \phi_j \right) \, dz. \label{eq:matrixN}
\ee
As $ \mathbf{N} $ is a Hermitian matrix, the matrix $ {\rm i} \mathbf{N} $ is skew-Hermitian, and since $ \mathbf{\Lambda} $ is Hermitian, the right hand side of equation (\ref{eq:Heisenberg}) corresponds to the decomposition of (\ref{eq:sys1}) into its Hermitian and skew-Hermitian part:
\be
\frac{\partial}{\partial t} \mathcal{L} w =
\underbrace{ \frac{1}{\Rey} \mathcal{L}^2 w
  + {\rm i} \alpha  \left(  D U D w + \frac{1}{2} D^2 U w \right)}_{Hermitian}
+ \underbrace{ {\rm i} \alpha  \left( \frac{1}{2} D^2 U w - D U D w - U \mathcal{L} w \right) }_{skew-Hermitian}.
\ee
Returning to equation (\ref{eq:Heisenberg}), matrix $ \mathbf{\Lambda} $ can be interpreted as a base Hamiltonian of the system, whereas $ - {\rm i} \mathbf{N} $ acts as a perturbation potential. In this view the base perturbations are growing or decaying exponential functions under action of a conservative potential. If we had chosen the eigenfunctions of (\ref{eq:eigenFrequency}), $ \mathbf{N} $ would be diagonal. In this picture $ {\rm i} \mathbf{N } $ would be the base Hamiltonian and $ \mathbf{\Lambda} $ a perturbation potential. The base perturbations in this case would be neutrally stable waves. This choice of discretization and the approximation by Feynman path integrals of the fundamental solution shall be discussed in more detail in appendix \ref{sec:pathIntegral}. For the above mentioned reasons, we shall only consider the eigenfunctions of (\ref{eq:davisPSE}) in the following. Equation (\ref{eq:Heisenberg}) can be slightly rewritten:
\be
\frac{d}{dt}\mathbf{c} = \mathbf{L} \mathbf{c} - {\rm i} \mathbf{M} \mathbf{c}. \label{eq:Heisenberg2}
\ee
The matrix $ \mathbf{L} $ is still diagonal but with its elements defined by
\be
L_{ij} = \left(\lambda_i - {\rm i} \omega_i\right) \delta_{ij},
\ee
where
\be
\omega_i = \frac{\alpha}{2k^2}
\int \limits_a^b \frac{1}{2} D^2 U | \phi_i |^2
+ U \left( | D\phi_i |^2 + k^2 | \phi_i |^2 \right) \, dz \label{eq:baseFrequency}
\ee
is the diagonal element of $ \mathbf{N} $. With $ \mathbf{L} $ as the base Hamiltonian, the base perturbations would be growing or decaying waves with frequency $ \omega_i $ given by (\ref{eq:baseFrequency}). Independent on the view adopted, the growth rate $\sigma$ of the resulting nonmodal Tollmien-Schlichting wave is given by:
\be
\sigma = \frac{1}{2} \frac{1}{E_w} \frac{d E_w}{dt} = \frac{1}{2}
\frac{ \frac{d \mathbf{c}^\dagger }{dt} \mathbf{c} + \mathbf{c}^\dagger \frac{d \mathbf{c} }{dt}   } { \mathbf{c}^\dagger \mathbf{c} }  = \frac{ \mathbf{c}^\dagger \mathbf{\Lambda} \mathbf{c}  } { \mathbf{c}^\dagger \mathbf{c} }.
\ee
Likewise the frequency $ \omega $ of the nonmodal Tollmien-Schlichting wave can be computed by:
\be
\omega = \frac{ \mathbf{c}^\dagger \mathbf{N} \mathbf{c}  } { \mathbf{c}^\dagger \mathbf{c} }.
\ee
The decompositions (\ref{eq:Heisenberg}) or (\ref{eq:Heisenberg2}) suggest a first approximation by neglecting the skew-Hermitian part, ie. the matrices $ \mathbf{N} $ or $ \mathbf{M} $, respectively. As the VKD-modes are uncoupled in this case, the optimal perturbation would be identical to the zeroth VKD-mode, displaying extreme growth for rather small Reynolds numbers. This contrasts with observations for many flows, such as Couette flow, which display only weak growth for two-dimensional perturbations. On the other hand, flows with adverse pressure gradients, such as the boundary layer flow under a solitary wave, have a tendency to favor two-dimensional perturbations. 

Equation (\ref{eq:Heisenberg}) provides a way to understand the effective growth of nonmodal Tollmien-Schlichting waves. As mentioned above, in order for growth of nonmodal Tollmien-Schlichting waves to be possible, at least the largest eigenvalue $ \lambda_0 $ needs to be positive. In order to simplify the following discussion, we shall assume that only $ \lambda_0 $ is positive and all other eigenvalues $ \lambda_i $, $ i = 1,\ldots $ are negative. In this case, the only VKD mode able to extract energy from the base flow is $ \phi_0 $. The first line of equation (\ref{eq:Heisenberg}) reads as follows:
\be
\frac{d}{dt} c_0 = \lambda_0 c_0 - {\rm i} \sum \limits_{j=0} N_{0j} c_j. \label{eq:firstComponent}
\ee
As $ \lambda_0 $ is the only positive eigenvalue, the first term  on the right hand side (\ref{eq:firstComponent}) is the only production term present in the system. The second term on the right hand side is a dispersion term distributing energy from the zeroth VKD mode to other VKD modes or vice versa. Therefore, even if the numeric value of $ \lambda_0 $ is large, dispersion between VKD modes can lead to a drain of energy towards other VKD modes, which stabilizes the system, as these modes dissipate energy. As we shall see in section \ref{sec:numerics}, the dispersion term in equation (\ref{eq:firstComponent}) is of principal importance for the evolution of nonmodal Tollmien-Schlichting waves. A bound for this term can be found by:
\bea 
| \sum_{j} N_{0j} c_j |^2 &=& | -\alpha 
\int \limits_a^b \frac{1}{4k^2} D^2 U \phi_0^\dagger w
+ \frac{1}{2k^2} U \left( D \phi_0^\dagger D w + k^2 \phi_0^\dagger w\right) \, dz |^2 \\
& \leq & \frac{\alpha^2}{4 k^4} | \int \limits_a^b  | \frac{1}{2} 
 D^2 U  \phi_0^\dagger w
- D (U  D \phi_0^\dagger ) w + k^2  U \phi_0^\dagger w| \, dz |^2  \\
& \leq & \frac{\alpha^2}{4 k^4} 
\int \limits_a^b  | \frac{1}{2} 
 D^2 U \phi_0^\dagger 
- D (U  D \phi_0^\dagger ) + k^2  U \phi_0^\dagger |^2 \, dz
\int \limits_a^b  | w |^2 \, dz  \\
&\leq& \frac{\alpha^2}{2 k^4} 
\int \limits_a^b  | \frac{1}{2} 
 D^2 U \phi_0^\dagger 
- D (U  D \phi_0^\dagger ) + k^2  U \phi_0^\dagger |^2 \, dz 
\times \nonumber \\
& & \quad \quad
 \frac{1}{2k^2} \int \limits_a^b  | D w|^2 + k^2| w |^2 \, dz.
\eea
This allows us to define a measure $ m_{0,\max} $ of the dispersion of the most dangerous VDK mode for the base flow $ U $ in question:
\be
m_{0,\max} = \max_{\mathbf{c}} \frac { \sum_{i,j} c_i^\dagger N_{i0} N_{0j} c_j }{ \sum_i c_i^\dagger c_i } 
\leq \frac{\alpha^2}{2 k^4} 
\int \limits_a^b  | \frac{1}{2} 
 D^2 U \phi_0 
- D (U  D \phi_0 ) + k^2  U \phi_0 |^2 \, dz  \label{eq:measure}
\ee
The left hand side of inequality (\ref{eq:measure}), ie. the measure $ m_{0,\max} $, can easily be found numerically by computing 
the spectral radius of the matrix with elements $ N_{i0}N_{0j} $. In principle, the analytic bound on the right hand side of (\ref{eq:measure}) gives us a possibility to relate the shape of the base flow profile $ U $ to its stability properties. Its interpretation is, however, not straightforward as $ \phi_0 $ implicitly is a function of $ U $, resulting from eigenvalue system (\ref{eq:davisPSE}). In principle, base flow profiles $ U $ with smaller bounds would be more unstable than those with larger bounds. A crude heuristic argument would suggest that profiles with oscillatory behavior around zero would probably be good candidates for unstable flows as a simple substitution of $ U $ by $ exp \pm {\rm i} \sqrt{2} k z $ reduces the argument of the integral to a single term. On the other hand, this heuristic argument would also imply that the perturbation with wavenumber $ k $ bringing $ D^2U/2 + k^2 U $ as close to zero as possible would experience larger growth. \\

Another estimate of the dispersion properties of the zeroth VDK mode is given by the temporal change of its energy, $ |c_0|^2 $. From equation (\ref{eq:firstComponent}), we obtain:
\bea
\frac{d}{dt} | c_0 |^2 &=& 2\lambda_0 | c_0 |^2 + 
{\rm i} \sum \limits_{j=0} c_0^\dagger M_{0k} c_j - 
c_j^\dagger M_{0j}^\dagger c_0 \\
&=&2\lambda_0 | c_0 |^2 + 
{\rm i} \sum \limits_{j=0} c_0^\dagger M_{0j} c_j - 
c_j^\dagger M_{j0} c_0 \\
&=& 2\lambda_0 | c_0 |^2 - 
2  \sum \limits_{j=0} {\rm imag} \left( c_0^\dagger M_{0j} c_j \right)
\eea
Besides the measure $ m_{0,\max} $, which is purely a property of the base flow $ U $, we shall investigate the following quantities $ m_0 $ and $ n_0 $, being actual dispersion measures for a given nonmodal Tollmien-Schlichting wave $ w $: 
\bea
m_0 &=& \frac { \sum_{i,j} c_i^\dagger N_{i0} N_{0j} c_j }{ \sum_i c_i^\dagger c_i } \\
&=& \frac{ \alpha^2}{ 4k^4 E_w} | 
\int \limits_a^b \frac{1}{2} D^2 U \phi_0^\dagger w
+ U \left( D \phi_0^\dagger D w + k^2 \phi_0^\dagger w\right) \, dz |^2 \label{eq:dispersionMeasureM0} \\
n_0 &=& {\rm i}\sum \limits_{j=0} c_0^\dagger M_{0j} c_j - 
c_j^\dagger M_{j0} c_0  \\
&=& \frac{{\rm i} \alpha}{ 2k^2}
\left( 
 \left\langle  \phi_0 , w \right\rangle \int \limits_a^b \frac{1}{2} D^2 U w^\dagger \phi_0 + U \left( D w^\dagger D \phi_0 + k^2 w^\dagger \phi_0 \right) \, dz  \right.
- \\
& & \quad \quad \quad \left. \left\langle  w, \phi_0 \right\rangle
\int \limits_a^b \frac{1}{2} D^2 U \phi_0^\dagger w + U \left( D \phi_0^\dagger D w + k^2 \phi_0^\dagger w \right) \, dz   \right) \label{eq:dispersionMeasureN0}
\eea
In the following, we shall continue the present investigation by means of a numerical analysis of three shear flows.

\subsection{Numerical results} \label{sec:numerics}

\subsubsection{Couette flow}

\begin{figure}
  \centering
    \includegraphics[width=\textwidth]{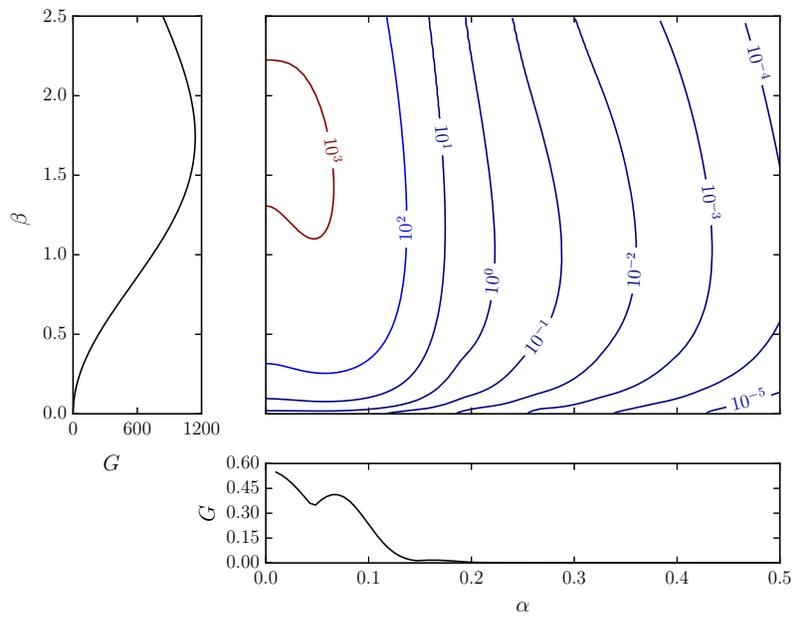}
    \caption{Isolines of $ G $ for the optimal perturbation for Couette flow at $ \Rey = 1000 $ at $ t = 117 $. The plots to the left and below the contour plot show a slice along the $ \beta $- and
$ \alpha $-axes, respectively.}
    \label{fig:Couette1}
\end{figure}

\begin{figure}
  \centering
    \includegraphics[width=\textwidth]{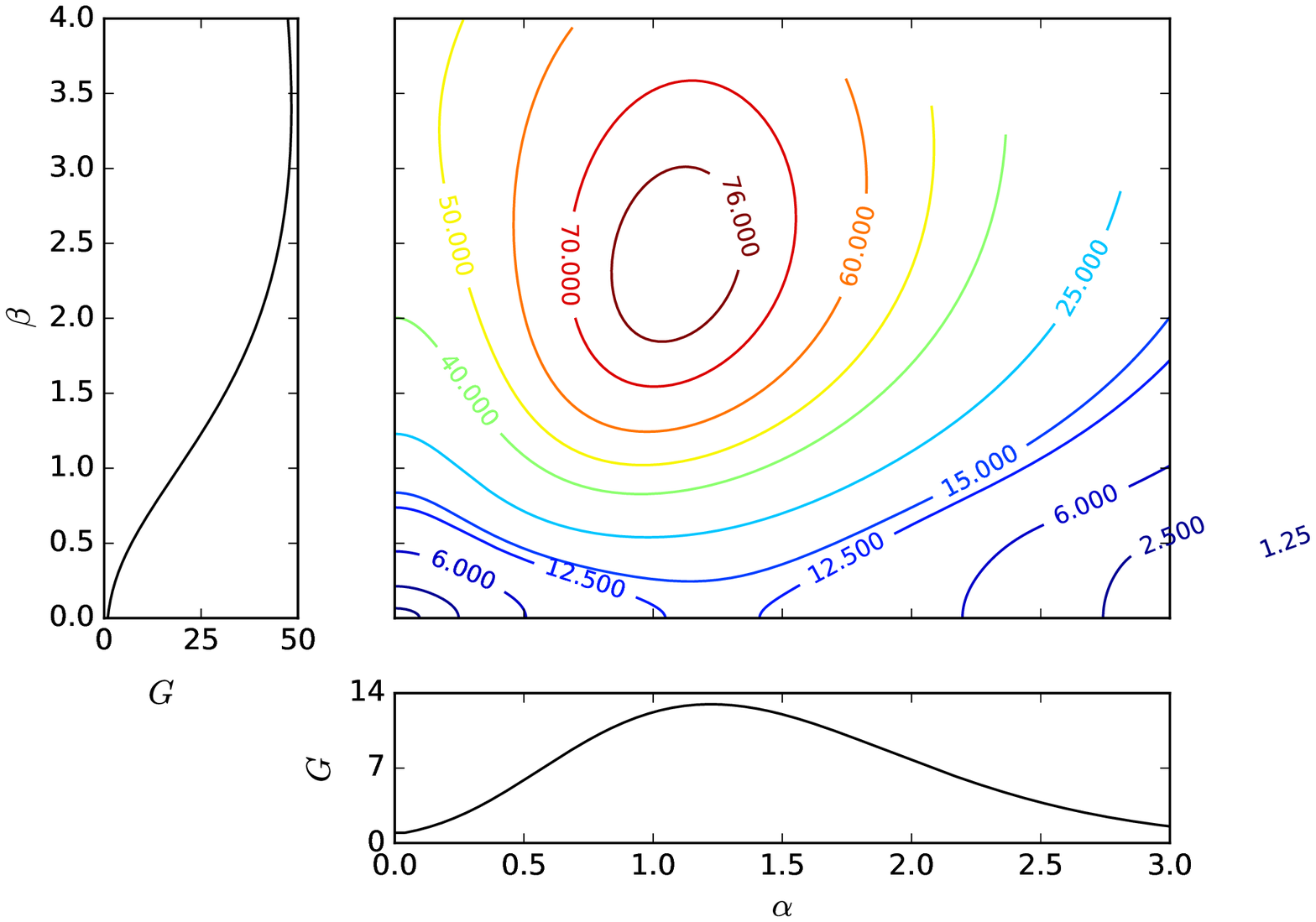}
    \caption{Isolines of $ G $ for the optimal perturbation for Couette flow at $ \Rey = 1000 $ at $ t = 8.7 $. The plots to the left and below the contour plot show a slice along the $ \beta $- and
$ \alpha $-axes, respectively.}
    \label{fig:Couette2}
\end{figure}

Optimal perturbations for Couette flow have been investigated by \cite{ButlerFarrell1992}. They found that for Couette flow at $ \Rey = 1000$, the global maximum is reached at $ t = 117 $ with $ G = 1184.6 $ for the optimal
perturbation with $ \alpha = 0.035 $ and $ \beta = 1.6 $. This result is plotted in figure \ref{fig:Couette1} by means of the present numerical solver, solving equations (\ref{eq:sys1}) and (\ref{eq:sys2}). In figure \ref{fig:Couette1} contour plots in Fourier space show that the largest amplifications are generated for streamwise streaks or superpositions close to being pure streamwise streaks. For $ \beta = 0 $, we observe that optimal nonmodal Tollmien-Schlichting waves display only decay. Perturbations with $ \beta = 0 $, have been found by \cite{ButlerFarrell1992} to display only weak growth. The maximum amplification reached by a perturbation with $ \beta = 0 $, they found, is at $ t = 8.7 $ with $ G = 13.0 $ for the optimal perturbation with $ \alpha = 1.21 $. However, as can be seen from figure \ref{fig:Couette2}, even in this case the perturbation with $ \alpha = 1.1 $ and $ \beta = 2.5 $ displays a much larger amplification which supports the conclusion by \cite{ButlerFarrell1992} that three dimensional perturbations and in particular streamwise streaks are dominant for Couette flow. Nevertheless, even if nonmodal Tollmien-Schlichting waves are subdominant to streamwise streaks, Couette flow is an illustrative example in order to understand the mechanics of nonmodal Tollmien-Schlichting waves. 

\begin{figure}
  \centering
    \includegraphics[width=\textwidth]{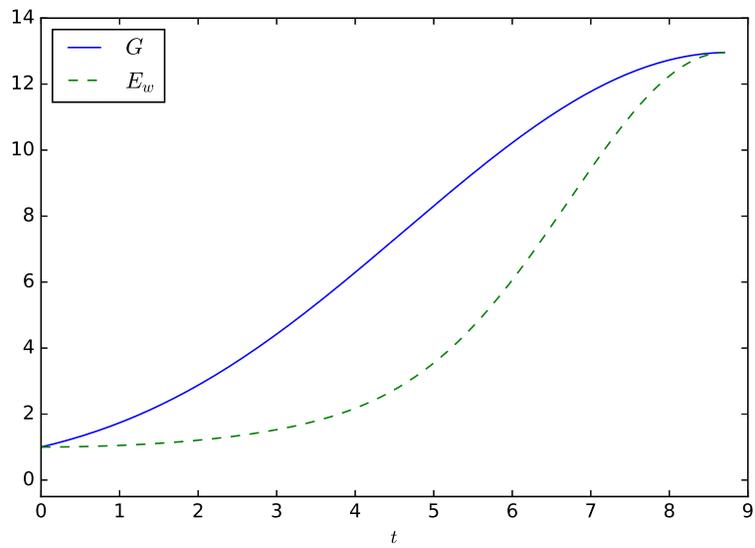}
    \caption{Amplification $ G $ of the optimal perturbation and temporal evolution of the energy $ E_w$ of the perturbation leading to a maximum amplification at $ t = 8.7 $ with $ \alpha = 1.21 $ and $ \beta = 0 $ for Couette flow at $ \Rey = 1000 $.}
    \label{fig:Couette3}
\end{figure}

\begin{figure}
  \centering
    \includegraphics[width=\textwidth]{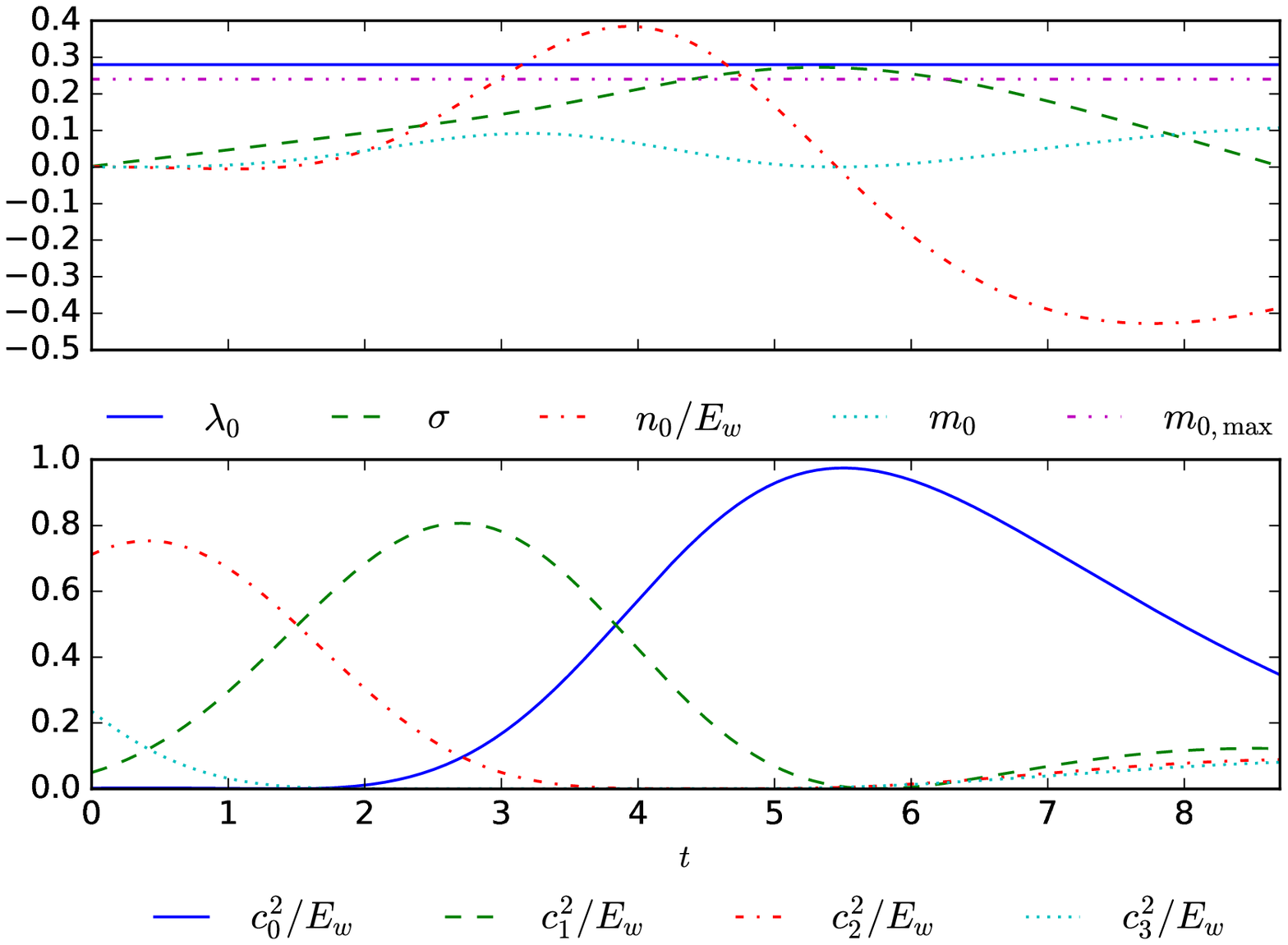}
    \caption{Temporal evolution of characteristic quantities of the nonmodal Tollmien-Schlichting wave with $ \alpha = 1.21 $, $ \beta = 0 $, for Couette flow at $ \Rey = 1000 $. Top: Growth rate $ \sigma $ with upper bound $ \lambda_0 $ and dispersion measures $ m_0 $, $ n_0 $ and $ m_{0,\max} $. Bottom: Energy contained in the first four VKD modes.}
    \label{fig:Couette4}
\end{figure}

\begin{figure}
  \centering
    \includegraphics[width=\textwidth]{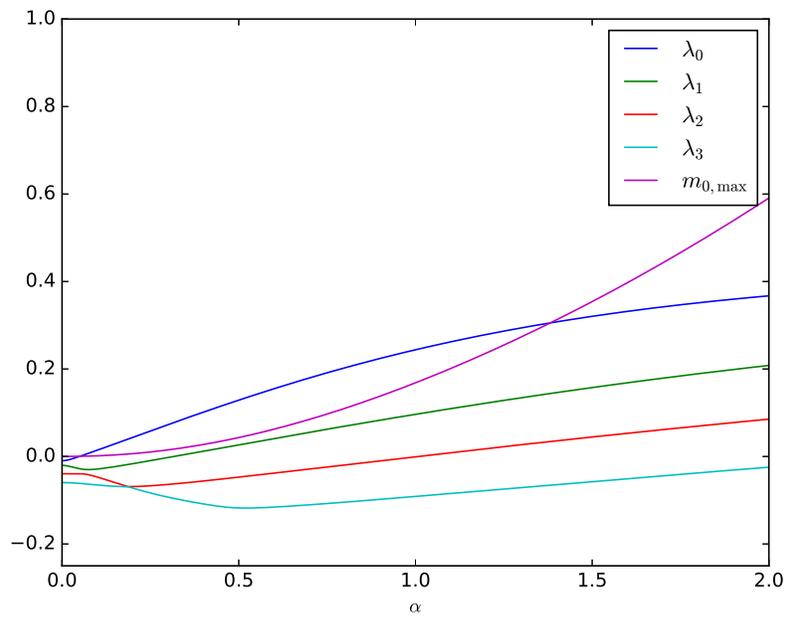}
    \caption{Growth rates $ \lambda_i $, equation (\ref{eq:eigenvalueLambda}), for Couette flow at $ \Rey = 1000 $ and dispersion measure $ m_{0,\max} $, equation (\ref{eq:measure}).}
    \label{fig:Couette5}
\end{figure}

In the following, we shall investigate the temporal evolution of nonmodal Tollmien-Schlichting waves. In particular, we concentrate on the nonmodal Tollmien-Schlichting wave with $ \alpha = 1.21 $ and $ \beta= 0 $ whose initial condition leads to the maximum at $ t = 8.7 $. The energy $ E_w $ of this Tollmien-Schlichting wave is plotted together with the amplification $ G $ of the optimal perturbation for $ \alpha = 1.21 $ and $ \beta = 0$ in figure \ref{fig:Couette3}. In the following, the initial energy $ E_w(t_0) $ at time $ t_0 $ is without loss taken to be unity. In figure \ref{fig:Couette4}, the growth rate $ \sigma $, equation (\ref{eq:growth}), of this nonmodal Tollmien-Schlichting wave is plotted. As can be observed the growth rate is bounded by $ \lambda_0 $, but comes close to this value at around $ t = 5.5 $. At approximately the same time the dispersion measure $ n_0 $, equation (\ref{eq:dispersionMeasureM0}) changes sign indicating that energy is first transferred from other VDK modes to the zeroth mode and then back again. This coincides well with the energy fraction contained in the zeroth VKD mode, see bottom figure \ref{fig:Couette4}. First $ n_0 $ is positive, allowing to accumulate energy in the zeroth VKD mode and to extract more energy from the base flow. However, as the energy fraction in the zeroth VKD mode grows, $ n_0 $ diminishes before crossing sign and inverting the energy transfer. In figure \ref{fig:Couette5}, the first four eigenvalues $ \lambda_i $ are plotted as a function of $ \alpha $. As mentioned above, in the present discussion the eigenvalues $ \lambda_i $ are enumerated in decreasing order. This implies that when plotting the eigenvalues as a function of, for example, the wave number $ \alpha $, the curves can display kinks as for the eigenvalues $ \lambda_2 $ and $ \lambda_3 $ at $ \alpha = 0.18 $, where a mode continuation approach would have suggested that the growth rate of one mode surpasses the one of another. 

For $ \alpha = 1.21 $, we observe that the first three VKD modes have a positive growth rate $ \lambda_i $. This fits the above picture, cf. figure \ref{fig:Couette4} bottom, where energy is transferred mainly between the first three growing VKD modes in order to obtain an optimal amplification at $ t = 8.7 $. If, we instead consider the optimal perturbation with maximum at $ t = 8.7 $ but with $ \alpha = 0.3 $, which, cf. figure \ref{fig:Couette5}, has only a single positive eigenvalue $ \lambda_0 $, we observe a somewhat different picture. The energy is mainly concentrated in the zeroth mode.

In both cases, we observe that the dispersion measure $ m_0 $ stays well below its
bound $ m_{0,\max} $. 

 Reaching optimal amplification at some point in time means thus to optimize the transfer of energy between VDK modes in order to maximize the energy fraction contained in the growing VKD modes.

\begin{figure}
  \centering
  \begin{subfigure}[b]{0.5\textwidth}
    \includegraphics[width=\textwidth]{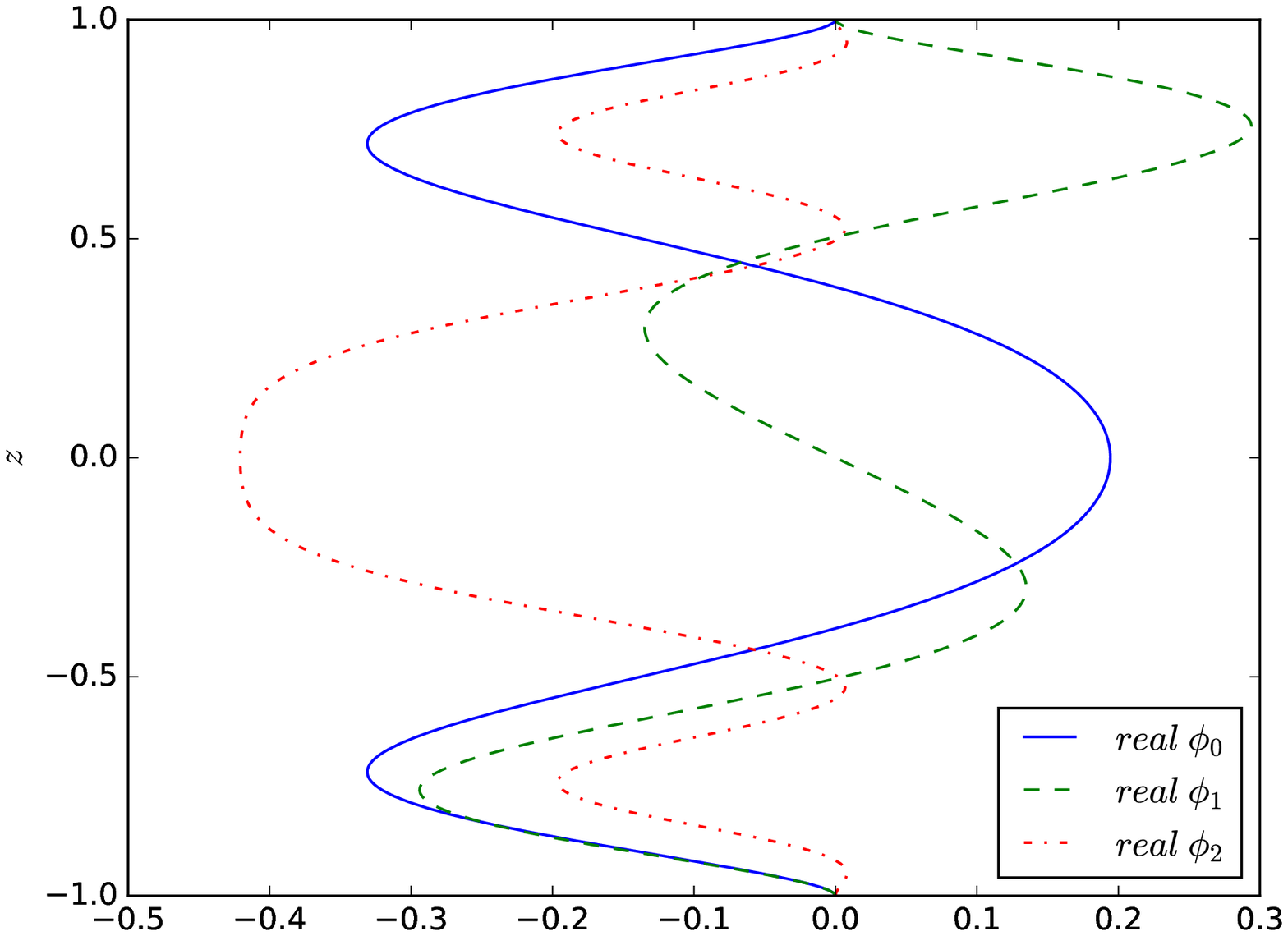}
\caption{Real part}
  \end{subfigure} \nolinebreak
  \begin{subfigure}[b]{0.5\textwidth}
    \includegraphics[width=\textwidth]{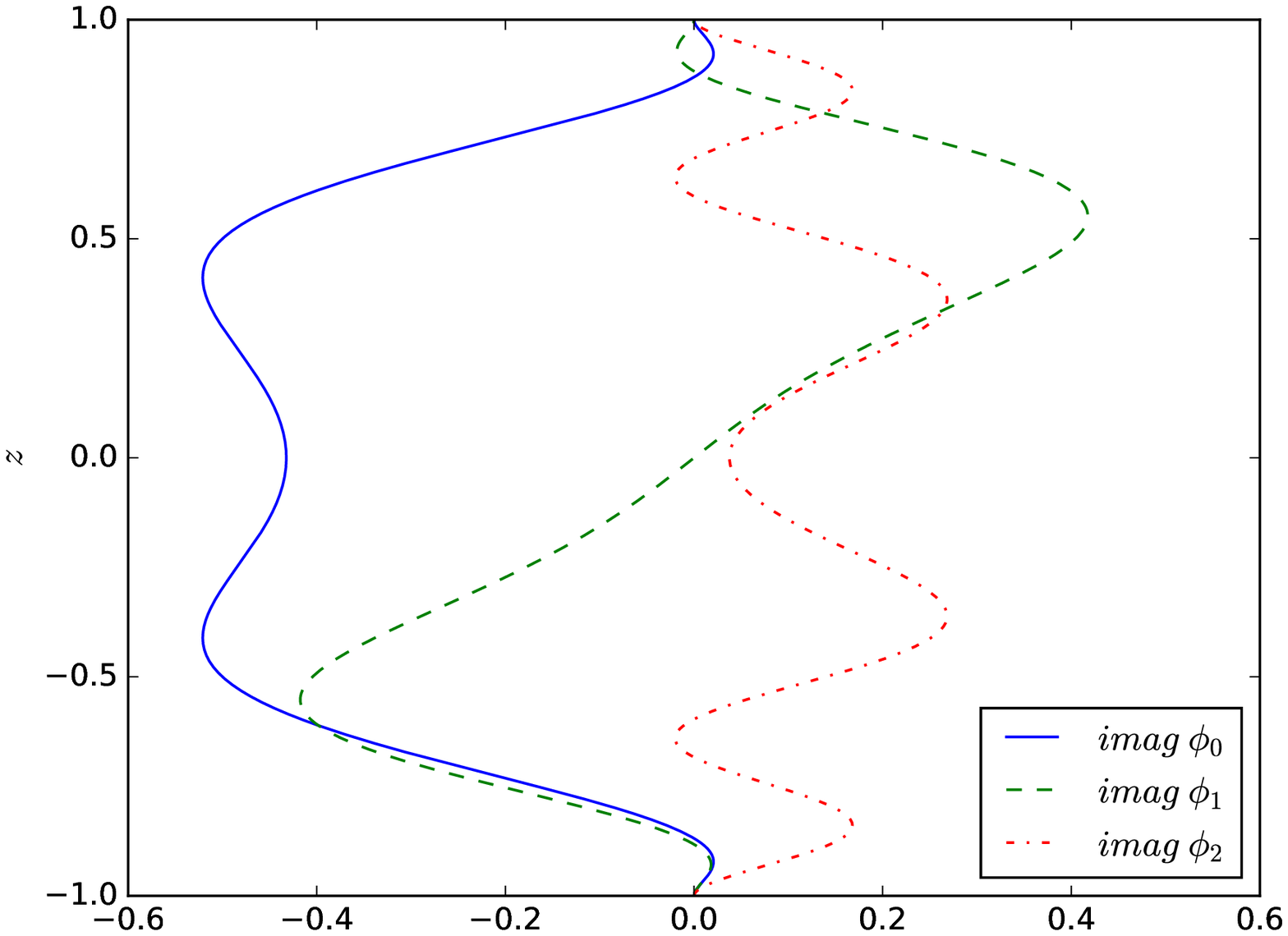}
    \caption{Imaginary part}
  \end{subfigure}
    \caption{Real and imaginary part of the first three VDK modes, with $ \alpha = 1.21 $ and $ \beta = 0$ for Couette flow at $ \Rey = 1000$.}
    \label{fig:Couette7}
\end{figure}

In figure \ref{fig:Couette7}, the real and imaginary part of the first three VKD modes with $ \alpha = 1.21 $ and $ \beta = 0 $ for Couette flow at $ \Rey = 1000 $ are plotted. The VKD modes are determined up to a constant phase, which in the present case is chosen such that the real part is symmetric around the origin, whereas the imaginary part is antisymmetric.

\begin{figure}
  \centering
    \includegraphics[width=\textwidth]{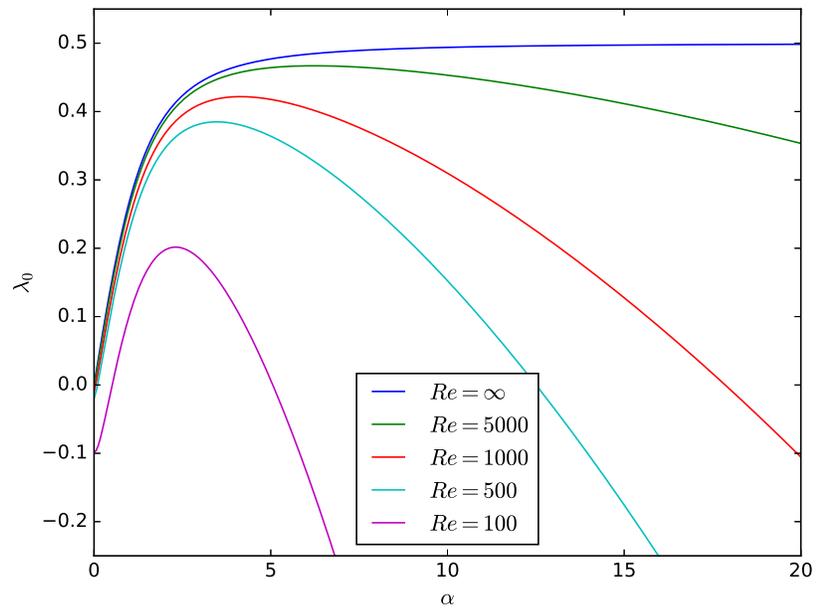}
    \caption{Growth rate $ \lambda_0 $ as a function of $ \alpha $ for Couette flow at different Reynolds numbers.}
    \label{fig:Couette8}
\end{figure}

Thanks to its simple form, we can compute some of the theoretical quantities analytically for Couette flow. The upper bound $ \lambda_{\max} $ on the growth rate, equation (\ref{eq:lambdaMax}), is for $ \beta = 0 $ equal to unity:
\be
\lambda_{\max} = 1.
\ee
Couette flow also allows to solve the eigenvalue equation (\ref{eq:inviscidDavis2}) for inviscid growth analytically. In fact, eigenvalue equation (\ref{eq:davisPSE}) for viscous growth is also solvable analytically. However, due to the complexity of the solution of a quartic equation, the analytic solution becomes cumbersome. The solution of (\ref{eq:inviscidDavis2}), can be written as follows:
\bea
\lambda_{n,\pm}^2 &=& \frac{ \alpha^2 }{ 4 \left( \alpha^2 + \gamma^2_n \right) },\\
\gamma_n &=& \frac{\pi}{2} ( 1 + n ), \\
r_n &=& \left\{ \begin{array}{cc} A_n \cos \gamma_n z & \mbox{for $ n $ even} \\
A_n \sin \gamma_n z & \mbox{for $ n $ odd} \end{array} \right.
\eea
We obtain thus for the zeroth VKD mode:
\bea
\lambda_0 &=& \frac{ \alpha }{ \sqrt{ 4 \alpha^2 + \pi^2 } } \\
\phi_0 &=& A_0 e^{ {\rm i} \frac{\alpha}{2 \lambda_0 } z} \cos \frac{ \pi}{2} z
\eea
Up to a phase, the constant $ A_0 $ is determined by the constraint that the energy of $ \phi_0 $ is unity:
\be
|A_0|^2 = \frac{ 4 \alpha^2}{ 4 \alpha^2 + \pi^2 } = 4 \lambda^2_0.
\ee
For large values of $ \alpha $, the inviscid growth rate $ \lambda_0 $ converges towards $ 1/2 $. On the other hand, as can be observed from figure \ref{fig:Couette8}, the growth rate $ \lambda_0 $ becomes negative for large values of $ \alpha $ in the viscous case. The larger the Reynolds number, the longer $ \lambda_0 $ stays positive. This is in accordance with formula (\ref{eq:growth}), where the viscous part scales as $ \alpha^2 $ for large values of $ \alpha $. This dissipation of small scales is absent in the inviscid case.

In the inviscid case, we can compute the inviscid dispersion bound analytically with $ \beta = 0$ for Couette flow:
\bea
\lefteqn{ \frac{1}{2\alpha^2}\int \limits_a^b  | \frac{1}{2} 
 D^2 U_\base \phi_0 
- D (U_\base  D \phi_0 ) + k^2 U_\base \phi_0 |^2 \, dz} \nonumber \\
& =& \frac{ 2 }{ 4 \alpha^2 + \pi^2} 
\left( \frac{\pi^2}{2} \left(2  + \frac{\pi^2}{3} \right) 
+ \alpha^2 \left( \pi^2 - 1 \right) +
4 
\alpha^4 \left( \frac{1}{3} - \frac{2}{\pi^2} \right) \right) \label{eq:analyticalDispersion}
\eea
From formula (\ref{eq:analyticalDispersion}), we observe that for $ \alpha \rightarrow \infty $, the bound displays a quadratic behavior with $ \alpha $. This behavior can also be seen for $ m_{0,\max} $ in the viscous case, cf. figure \ref{fig:Couette5}. As such $ \lambda_0 $ is for most Reynolds numbers growing faster for smaller $ \alpha $ than $ m_{0,\max} $, before $ m_{0,\max} $ overtakes $ \lambda_0 $, cf figure \ref{fig:Couette5}, indicating that for a certain value of $ \alpha $, growth and dispersion to other VKD modes reach break even. It is thus rather the quadratic growth of $ m_{0,\max} $ than the viscous dissipation of small scales, which is at the origin of a finite value of $ \alpha $ for which the optimal perturbation reaches a maximum amplification, for example the value $ \alpha = 1.21 $ for Couette flow at $ \Rey = 1000$.

\begin{figure}
  \centering
    \includegraphics[width=\textwidth]{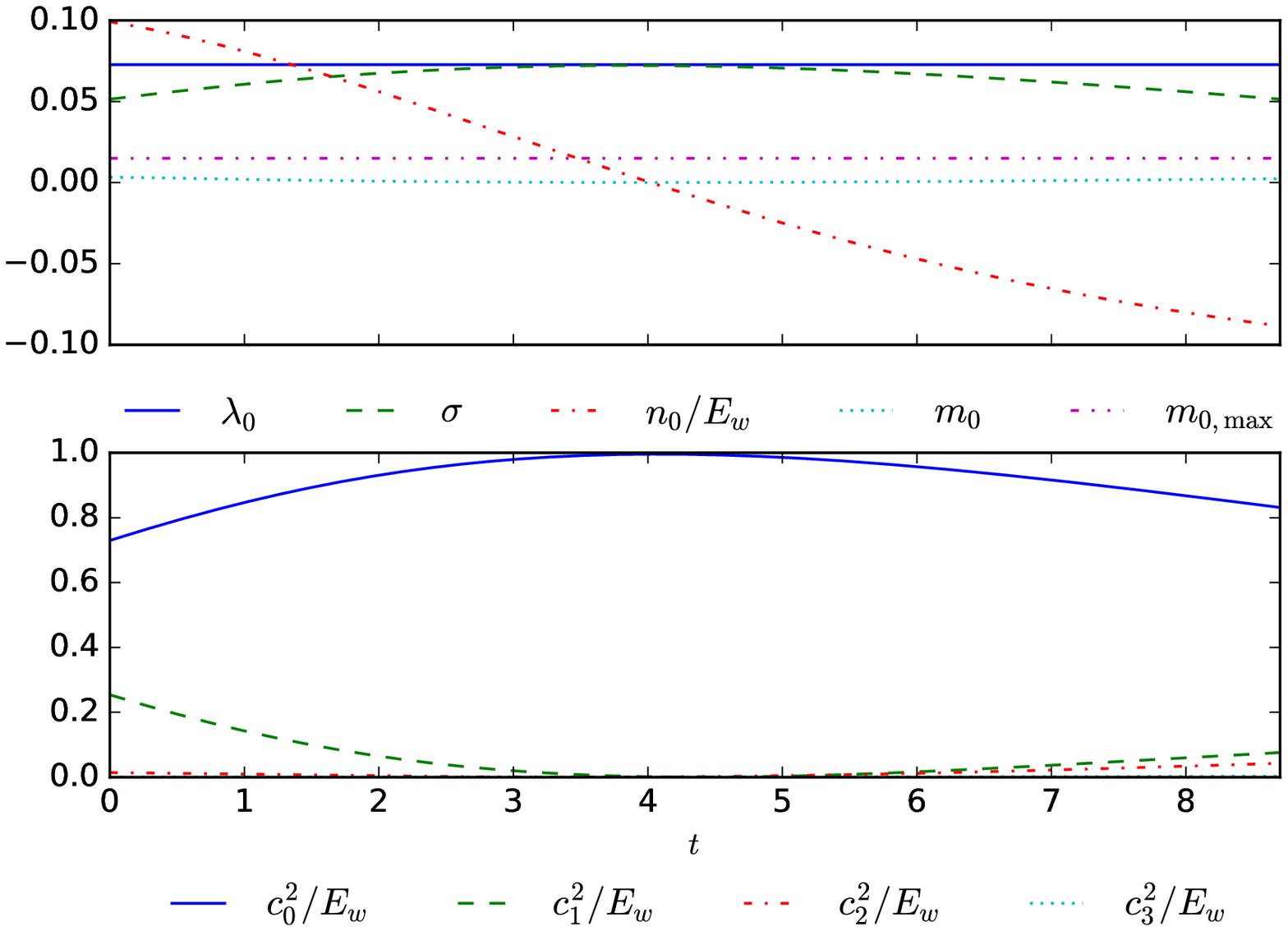}
    \caption{Temporal evolution of characteristic quantities of the nonmodal Tollmien-Schlichting wave with $ \alpha = 0.3 $, $ \beta = 0 $, for Couette flow at $ \Rey = 1000 $. Top: Growth rate $ \sigma $ with upper bound $ \lambda_0 $ and dispersion measures $ m_0 $, $ n_0 $ and $ m_{0,\max} $. Bottom: Energy contained in the first four VKD modes.}
    \label{fig:Couette6}
\end{figure}

\clearpage

\subsection{Poiseuille flow}

\begin{figure}
  \centering
    \includegraphics[width=\textwidth]{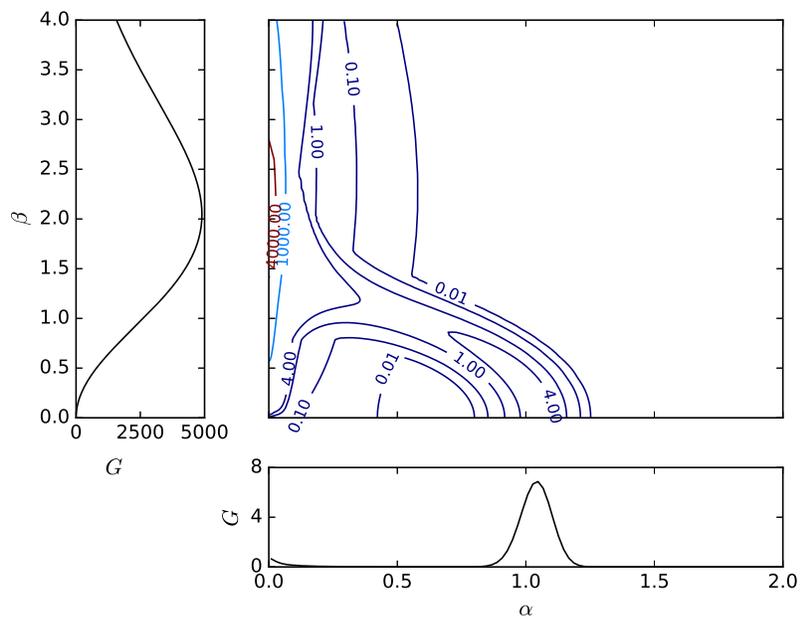}
    \caption{Isolines of $ G $ for the optimal perturbation for Poiseuille flow at $ \Rey = 5000 $ at $ t = 379 $. The plots to the left and below the contour plot show a slice along the $ \beta $- and
$ \alpha $-axes, respectively.}
    \label{fig:Poiseuille1}
\end{figure}

\begin{figure}
  \centering
    \includegraphics[width=\textwidth]{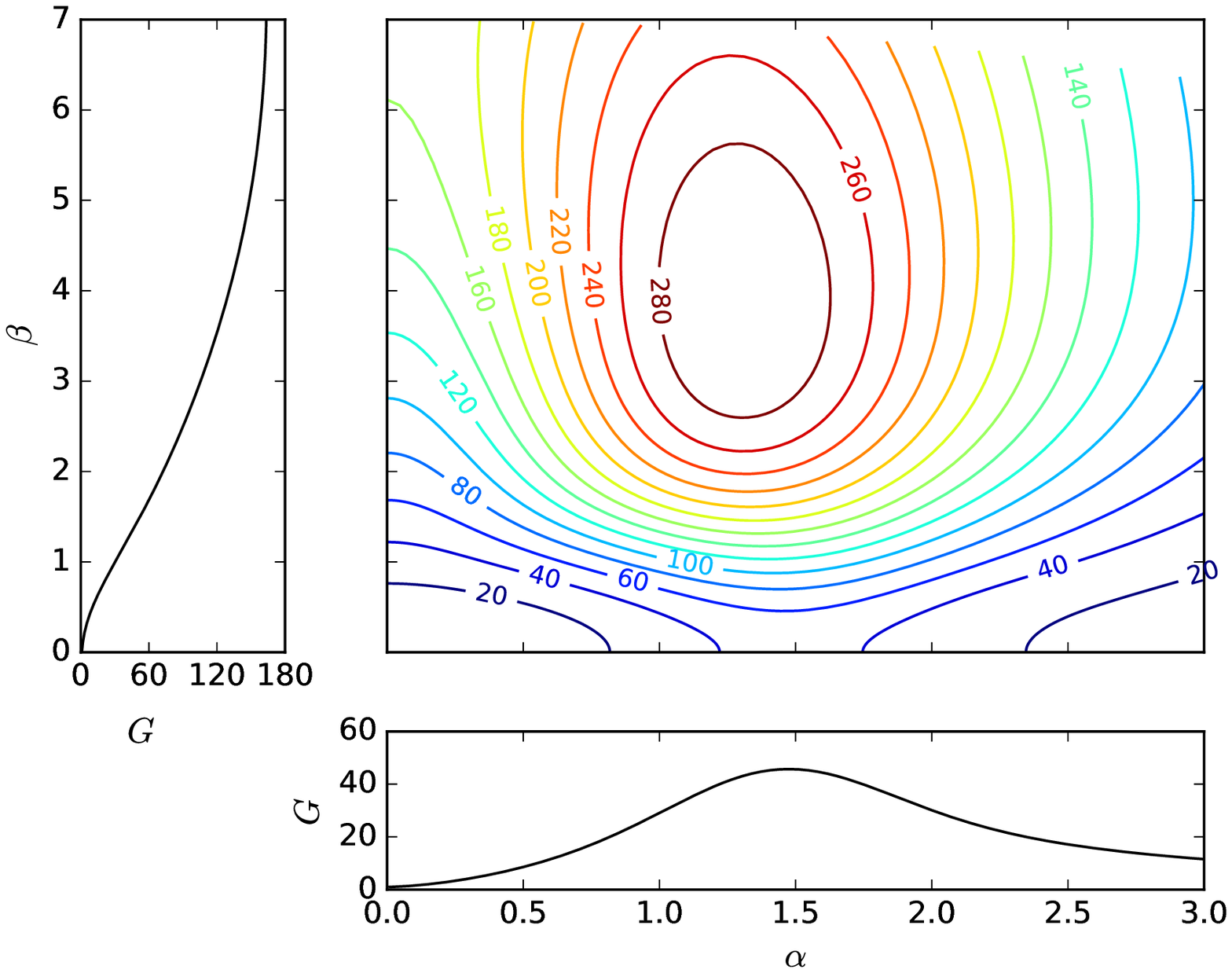}
    \caption{Isolines of $ G $ for the optimal perturbation for Poiseuille flow at $ \Rey = 5000 $ at $ t = 14.1 $. The plots to the left and below the contour plot show a slice along the $ \beta $- and
$ \alpha $-axes, respectively.}
    \label{fig:Poiseuille2}
\end{figure}

\begin{figure}
  \centering
    \includegraphics[width=\textwidth]{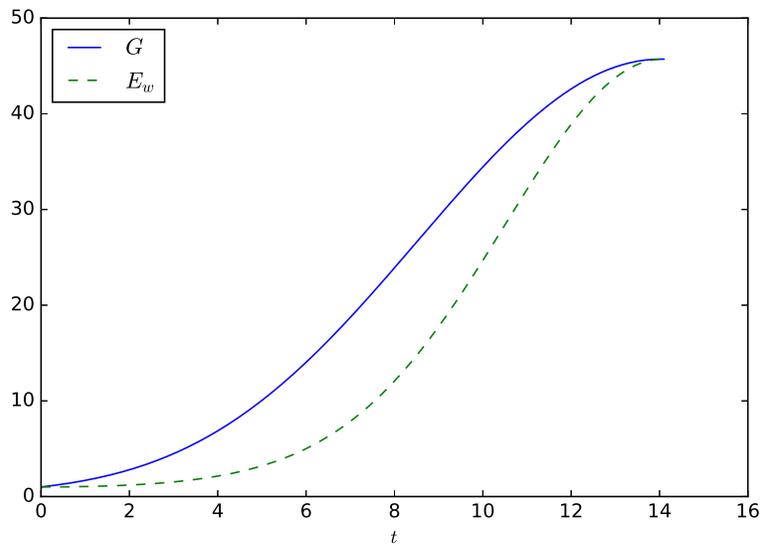}
    \caption{Amplification $ G $ of the optimal perturbation and temporal evolution of the amplification $ E_w $ of the perturbation leading to a maximum amplification at $ t = 14.1 $ with $ \alpha = 1.48 $ and $ \beta = 0 $ for Poiseuille flow at $ \Rey = 5000 $.}
    \label{fig:Poiseuille3}
\end{figure}

\begin{figure}
  \centering
  \includegraphics[width=\textwidth]{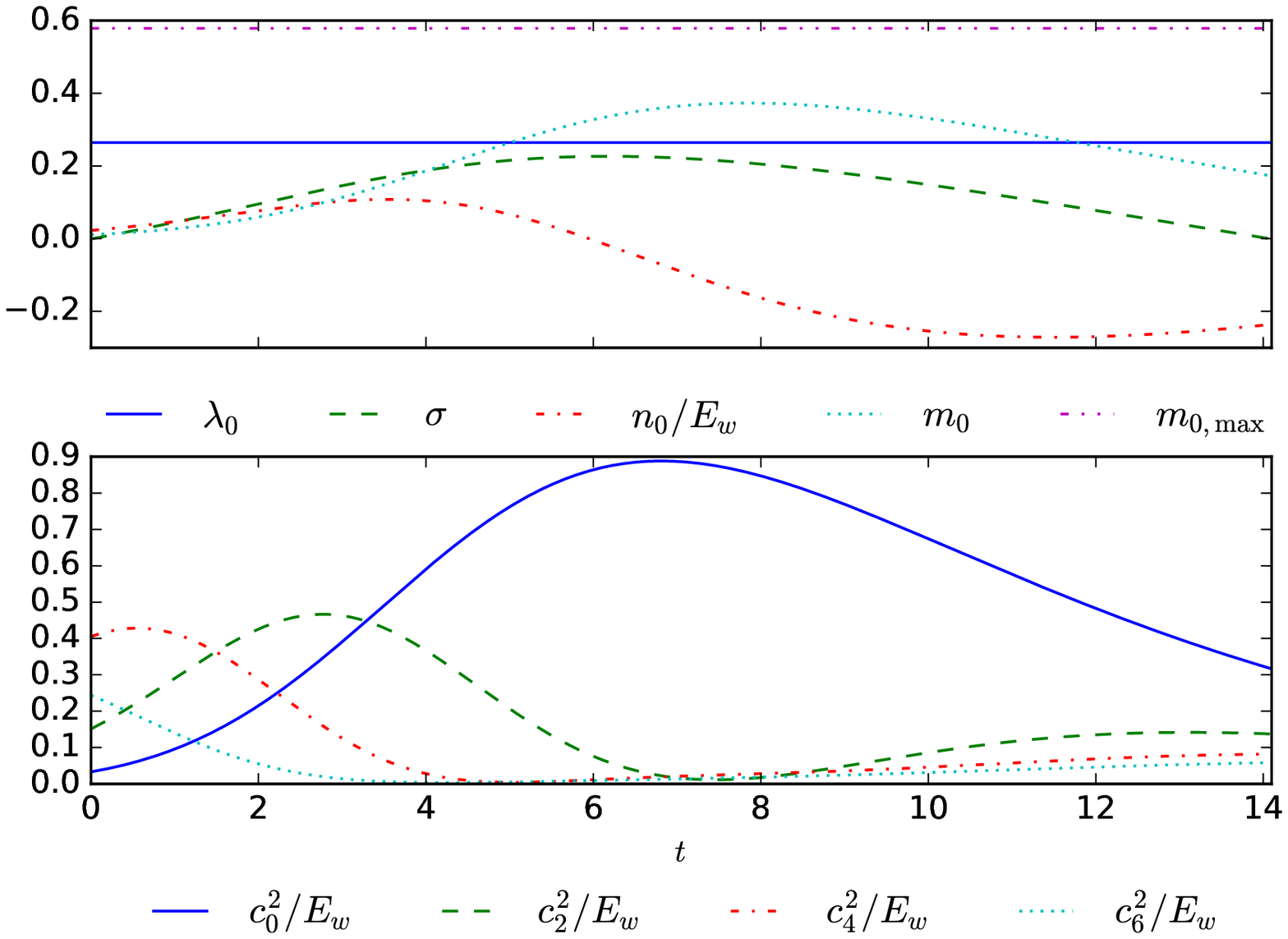}
    \caption{Temporal evolution of characteristic quantities of the nonmodal Tollmien-Schlichting wave with $ \alpha = 1.48 $, $ \beta = 0 $, for Poiseuille flow at $ \Rey = 5000 $. Top: Growth rate $ \sigma $ with upper bound $ \lambda_0 $ and dispersion measures $ m_0 $, $ n_0 $ and $ m_{0,\max} $. Bottom: Energy contained in even VKD modes.}
    \label{fig:Poiseuille4}
\end{figure}

\begin{figure}
  \centering
  \begin{subfigure}[b]{0.5\textwidth}
    \includegraphics[width=\textwidth]{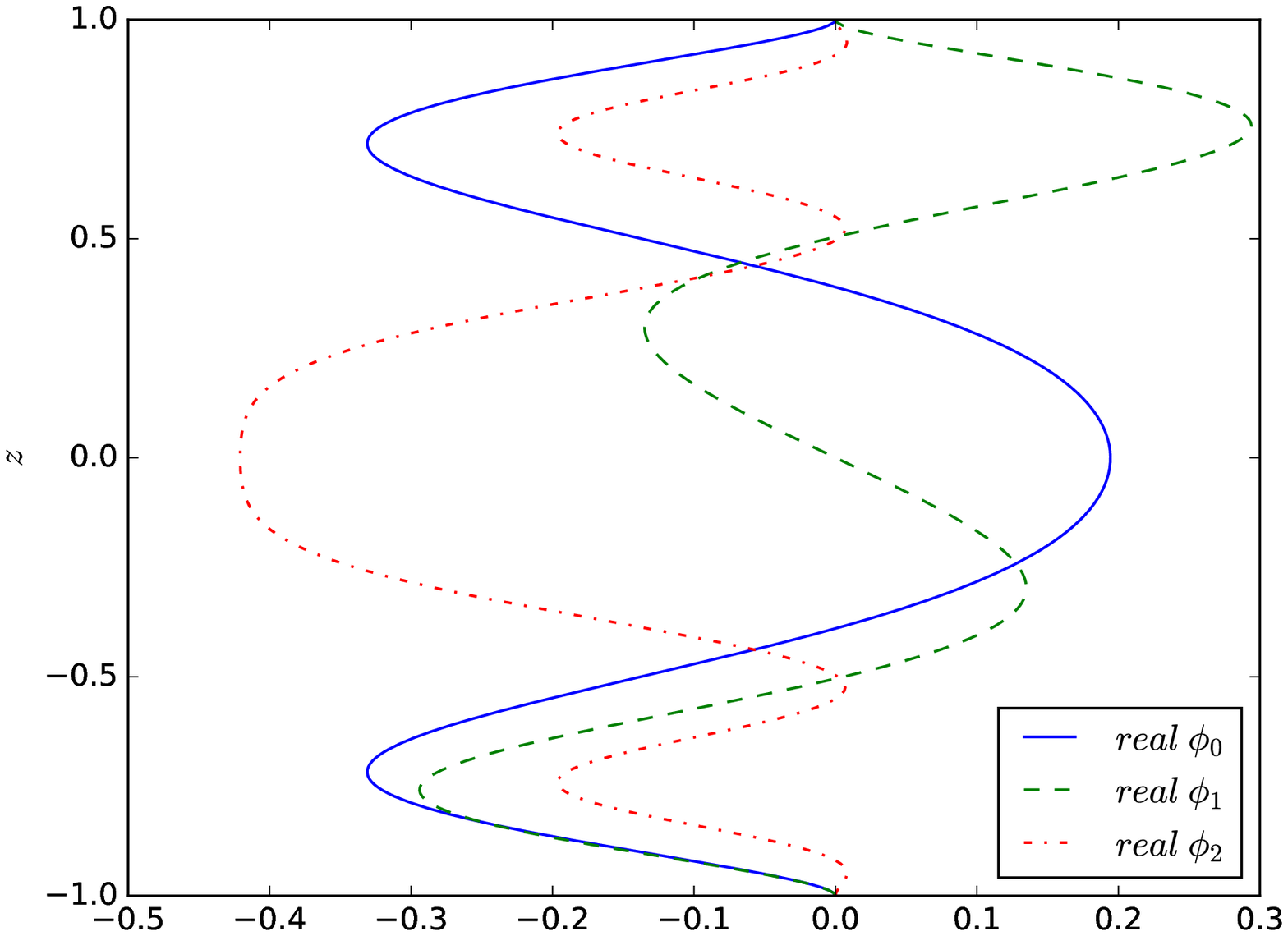}
\caption{Real part}
  \end{subfigure} \nolinebreak
  \begin{subfigure}[b]{0.5\textwidth}
    \includegraphics[width=\textwidth]{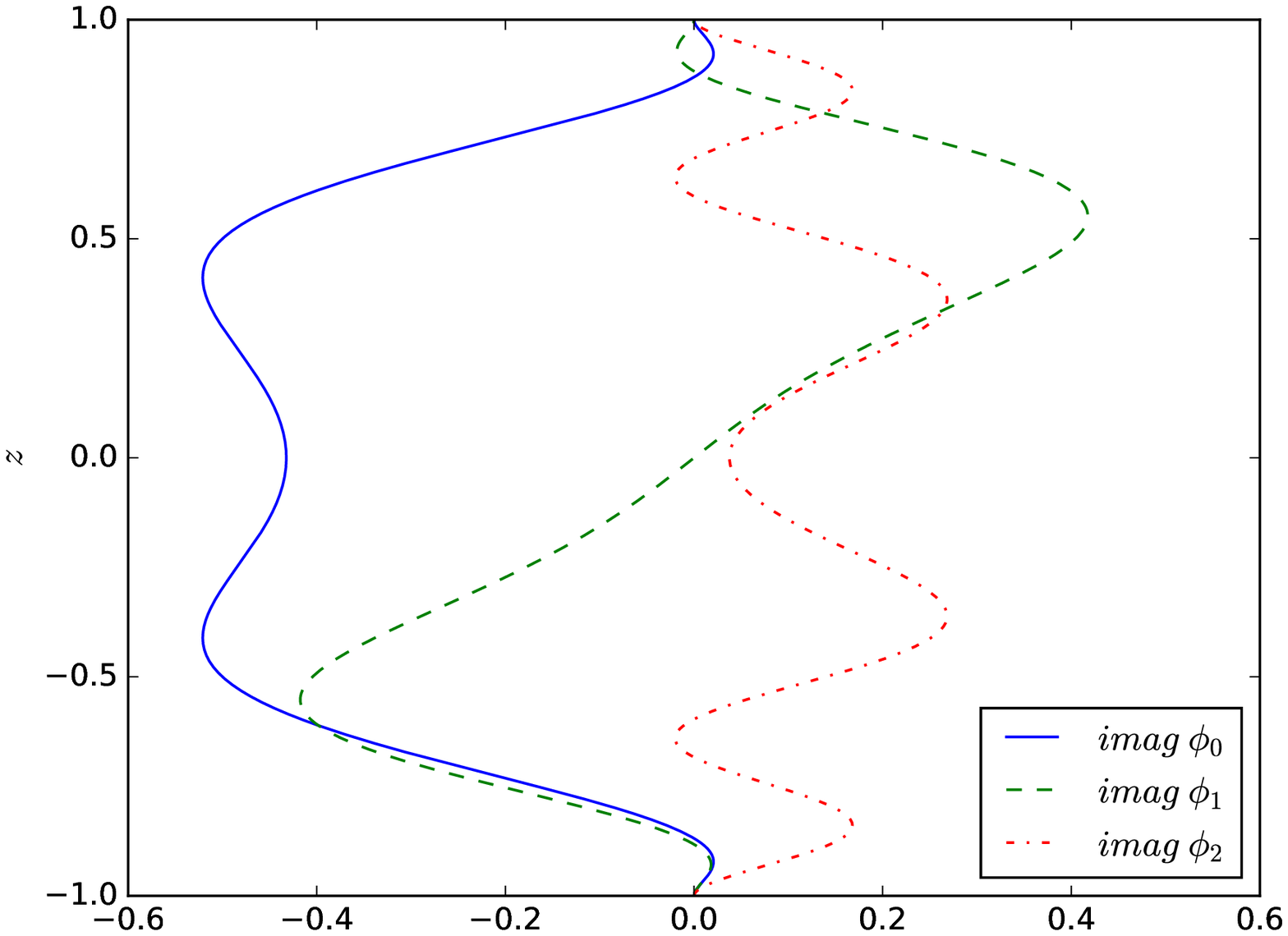}
    \caption{Imaginary part}
  \end{subfigure}
    \caption{Real and imaginary part of the first three VKD modes with $ \alpha = 1.48 $ and $ \beta = 0 $ for Poiseuille flow at $ \Rey = 5000$.}
    \label{fig:Poiseuille7}
\end{figure}

\begin{figure}
  \centering
    \includegraphics[width=\textwidth]{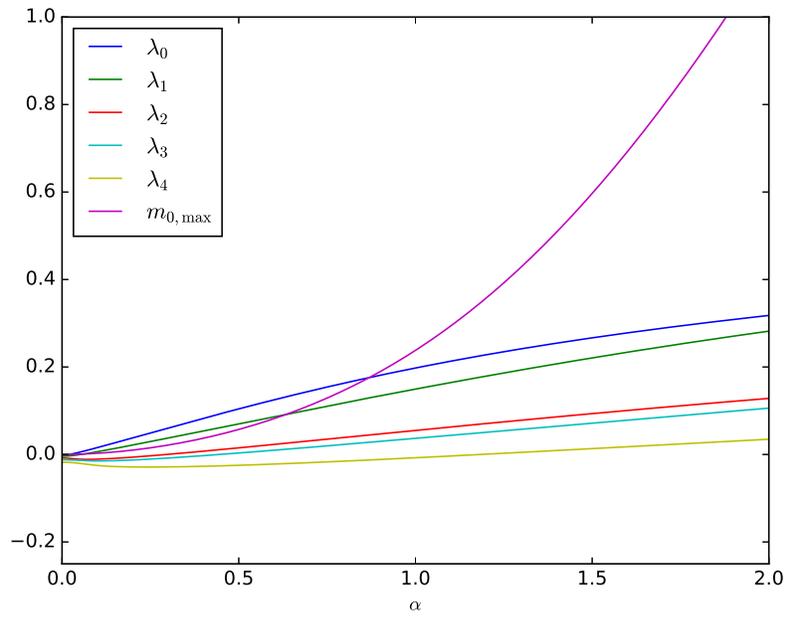}
    \caption{Growth rates $ \lambda_i $, equation (\ref{eq:eigenvalueLambda}), for Poiseuille flow at $ \Rey = 5000 $ and dispersion measure $ m_{0,\max} $, equation (\ref{eq:measure}).}
    \label{fig:Poiseuille5}
\end{figure}

\begin{figure}
  \centering
    \includegraphics[width=\textwidth]{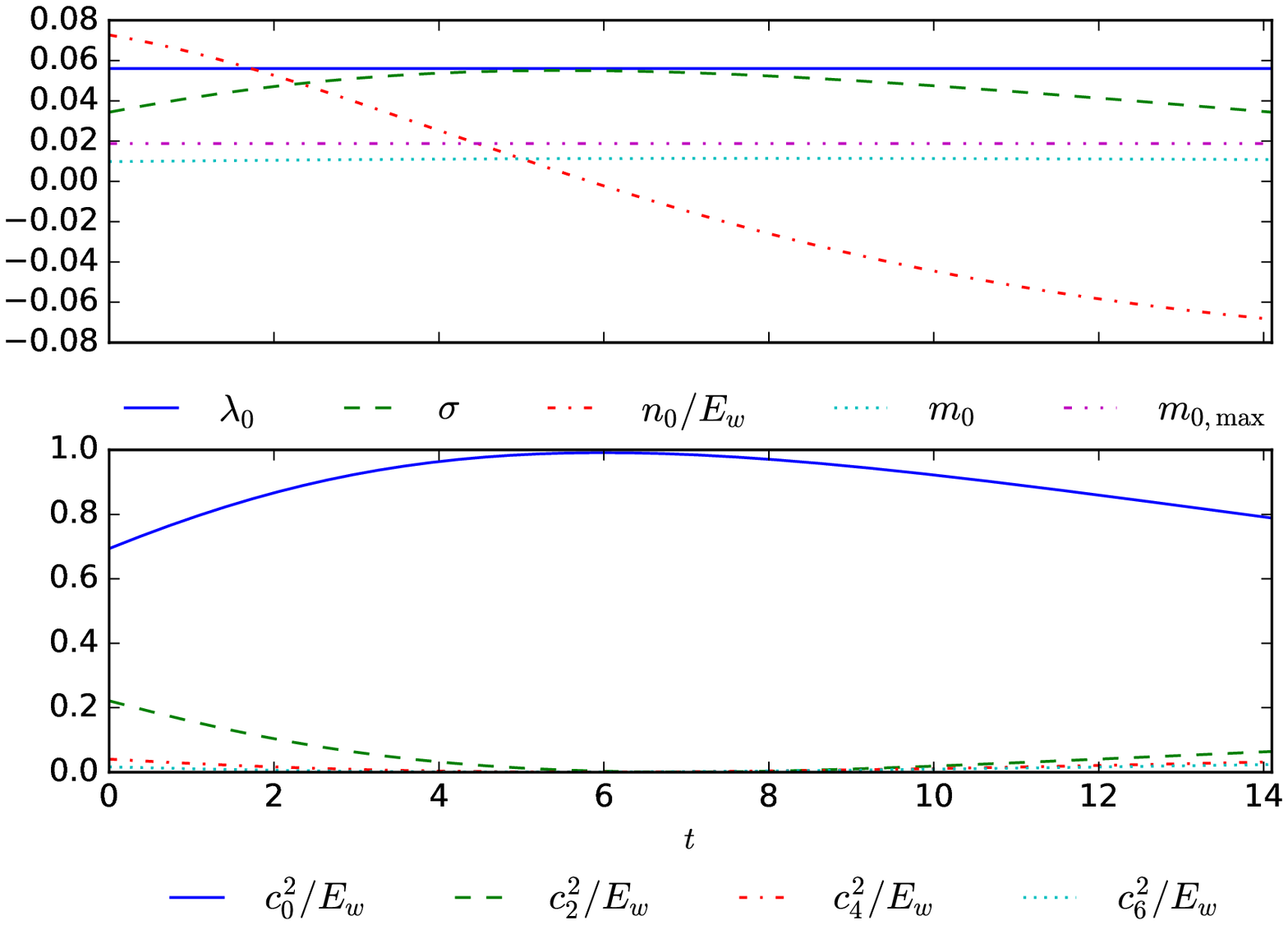}
    \caption{Temporal evolution of characteristic quantities of the nonmodal Tollmien-Schlichting wave with $ \alpha = 0.28 $, $ \beta = 0 $, for Poiseuille flow at $ \Rey = 5000 $. Top: Growth rate $ \sigma $ with upper bound $ \lambda_0 $ and dispersion measures $ m_0 $, $ n_0 $ and $ m_{0,\max} $. Bottom: Energy contained in even VKD modes.}
    \label{fig:Poiseuille6}
\end{figure}

Similar to Couette flow, \cite{ButlerFarrell1992} found that streamwise streaks are the dominant perturbations for Poiseuille flow. At a Reynolds number of $ \Rey = 5000 $, \cite{ButlerFarrell1992} calculated that the streamwise streak with $ \alpha = 0 $ and $ \beta = 2.044 $ reaches the global maximum at $ t = 379 $, with $ G = 4897 $. In figure \ref{fig:Poiseuille1}, the amplification $ G $ of the optimal perturbation at $ t = 379 $ is plotted in Fourier space for Poiseuille flow at $ \Rey = 5000 $. As can be observed from this plot, streamwise streaks dominate over two-dimensional perturbations which display just a small peak around $ \alpha \approx 1.05 $. According to \cite{ButlerFarrell1992}, the global maximum for a two-dimensional perturbation ($ \beta = 0 $), is reached for the nonmodal Tollmien-Schlichting wave with $ \alpha = 1.48 $ at time $ t = 14.1 $. As for Couette flow, this maximum is only a saddle point in Fourier space, cf. figure \ref{fig:Poiseuille2}. The maximum at this time is reached by the superposition with $ \alpha \approx 1.4 $ and $ \beta \approx 4 $. As before, we concentrate on the evolution of the nonmodal Tollmien-Schlichting wave with $ \alpha = 1.48 $ and $ \beta = 0 $, being the optimal two-dimensional perturbation at $ t = 14.1 $, cf. figure \ref{fig:Poiseuille3}. When plotting the evolution of $ n_0 $, cf. figure \ref{fig:Poiseuille4}, we observe, as before that first energy is transferred to the zeroth VKD mode before being returned to other modes again. There is, however, a difference to Couette flow. The energy contained in VKD modes with an odd index is zero (not shown). When plotting the real and imaginary part of the first VKD modes, cf. figure \ref{fig:Poiseuille7}, we observe that VKD modes with even indices are even functions and VKD modes with odd indices are odd functions. From equation (\ref{eq:matrixN}), we can readily infer, that in this case, the matrix $ \mathbf{N} $ is banded, since we have for the elements of $ \mathbf{N} $:
\be
N_{ij} = 0 \quad \quad \mbox{if $ i+j $ odd }.
\ee
For this reason, only the even indexed VKD modes are part of the optimal perturbation, as no energy transfer between odd and even indexed VKD modes is possible. As for Couette flow, we observe a quadratic behavior of the dispersion measure $ m_{0,\max} $, cf. figure \ref{fig:Poiseuille5}. In addition, as we can infer from figure \ref{fig:Poiseuille5}, there are three VKD-modes with positive growth rates mainly transferring energy between each other as depicted in figure \ref{fig:Poiseuille4}. When choosing the value $ \alpha  = 0.28 $, the only positive growth rate is $ \lambda_0 $. As visible from figure \ref{fig:Poiseuille6}, we observe, similar to Couette flow, that most of the energy is contained in the zeroth mode. 

\begin{figure}
  \centering
    \includegraphics[width=\textwidth]{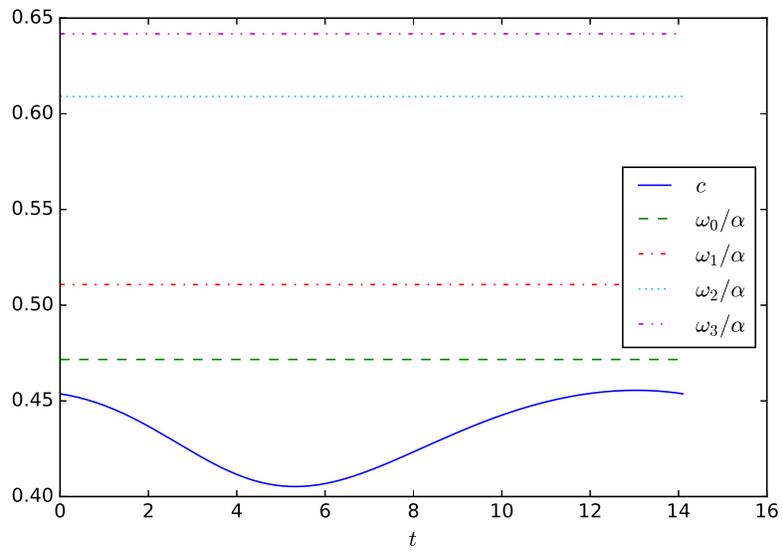}
    \caption{Phase speed $ c $ of the nonmodal Tollmien-Schlichting wave with $ \alpha = 1.48 $ and $ \beta = 0 $ and phase speed of the four first VDK modes, for Poiseuille flow at $ \Rey = 5000 $. }
  \label{fig:Poiseuille8}
\end{figure}

For the above Couette flow, the phase speed of the nonmodal Tollmien-Schlichting wave and the phase speed of the first VKD modes is zero (not shown). This is, however, not the case for Poiseuille flow at $ \Rey = 5000$, cf. figure \ref{fig:Poiseuille8}. The VKD modes travel at different phase speeds, equation (\ref{eq:baseFrequency}). The nonmodal Tollmien-Schlichting wave propagates with a phase speed close to the average velocity of the base flow.

\clearpage

\subsection{Boundary layer under a solitary wave}

\begin{figure} 
    \includegraphics[width=\textwidth]{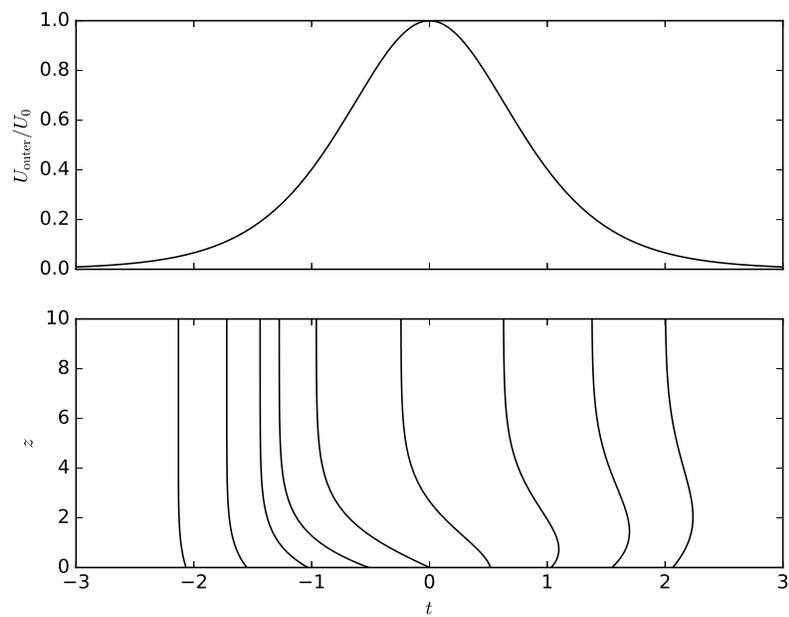}
    \caption{Inviscid outer flow $ U_\free $ 
and profiles of the horizontal velocity component in the boundary layer under a solitary wave moving from right to left. 
The profiles 
have been multiplied by 40.  
The value at $ z = 0 $ of the profiles shown corresponds to the
point in time $ t $, at which the profile has been taken.
The horizontal velocity
vanishes at $ z = 0 $ in order to satisfy the no-slip boundary condition.}
\label{fig:soliton1}
\end{figure}

Compared to the two preceding flow examples, the boundary layer flow under a solitary wave has significant distinct features. It is not only time dependent and defined in a semi-infinite interval, but most importantly, it develops an adverse pressure gradient for time $ t > 0 $. The flow is an idealization of the boundary layer flow under a solitary wave and was originally proposed as a model flow in \cite{SumerJensenSorensenFredsoeLiuCarstensen2010}. The outer flow of this boundary layer is given by:
\be
U_{outer}(2t/\Rey) =  {\rm sech}^2 \left( 2t/\Rey \right). \label{eq:solitonFormula}
\ee
Opposed to the present discussion, we remark that in \cite{VerschaevePedersenTropea2018}, time $ t $ was measured in the scale of the outer flow. 
As in \cite{VerschaevePedersenTropea2018}, the boundary layer equations are solved numerically to obtain the base flow $ U(2t/\Rey,z) $. In figure \ref{fig:soliton1}, the outer flow $ U_{outer} $ and some velocity profiles for selected times are plotted. For times $ t < 0 $ the flow accelerates, whereas for $ t > 0 $ it decelerates. The return flow developing due to the adverse pressure gradient for times $ t > 0 $ is clearly visible.

\begin{figure}
  \centering
  \begin{subfigure}[b]{0.5\textwidth}
    \includegraphics[width=\textwidth]{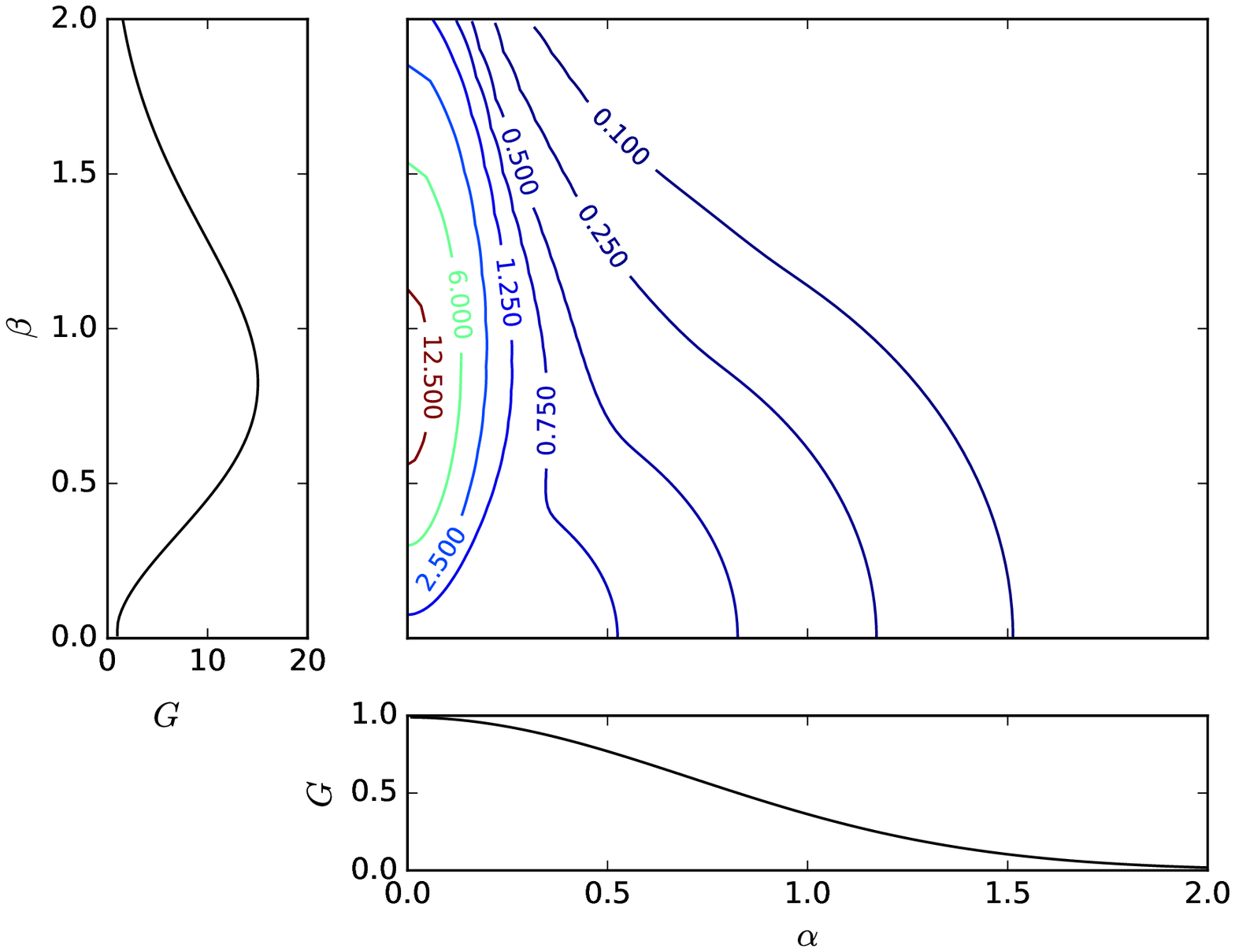}
    \caption{ $ 2t_0/\Rey = -1 $, $ 2t_1/\Rey = 0 $ }
    \label{fig:contourA}
  \end{subfigure} \nolinebreak
  \begin{subfigure}[b]{0.5\textwidth}
    \includegraphics[width=\textwidth]{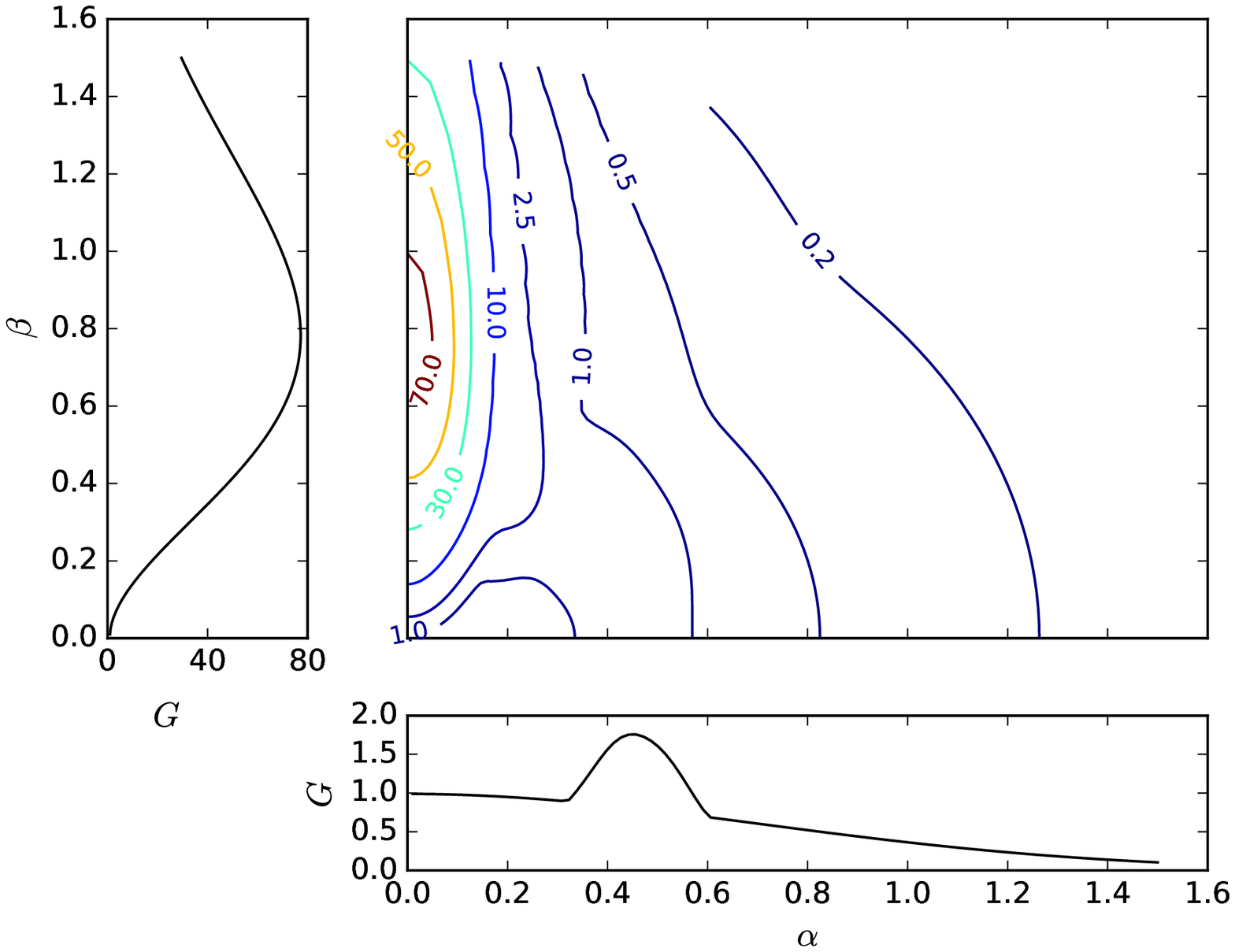}
    \caption{ $ 2t_0/\Rey = 0 $, $ 2t_1/\Rey = 1 $ }
    \label{fig:contourB}
  \end{subfigure}
  \begin{subfigure}[b]{0.5\textwidth}
    \includegraphics[width=\textwidth]{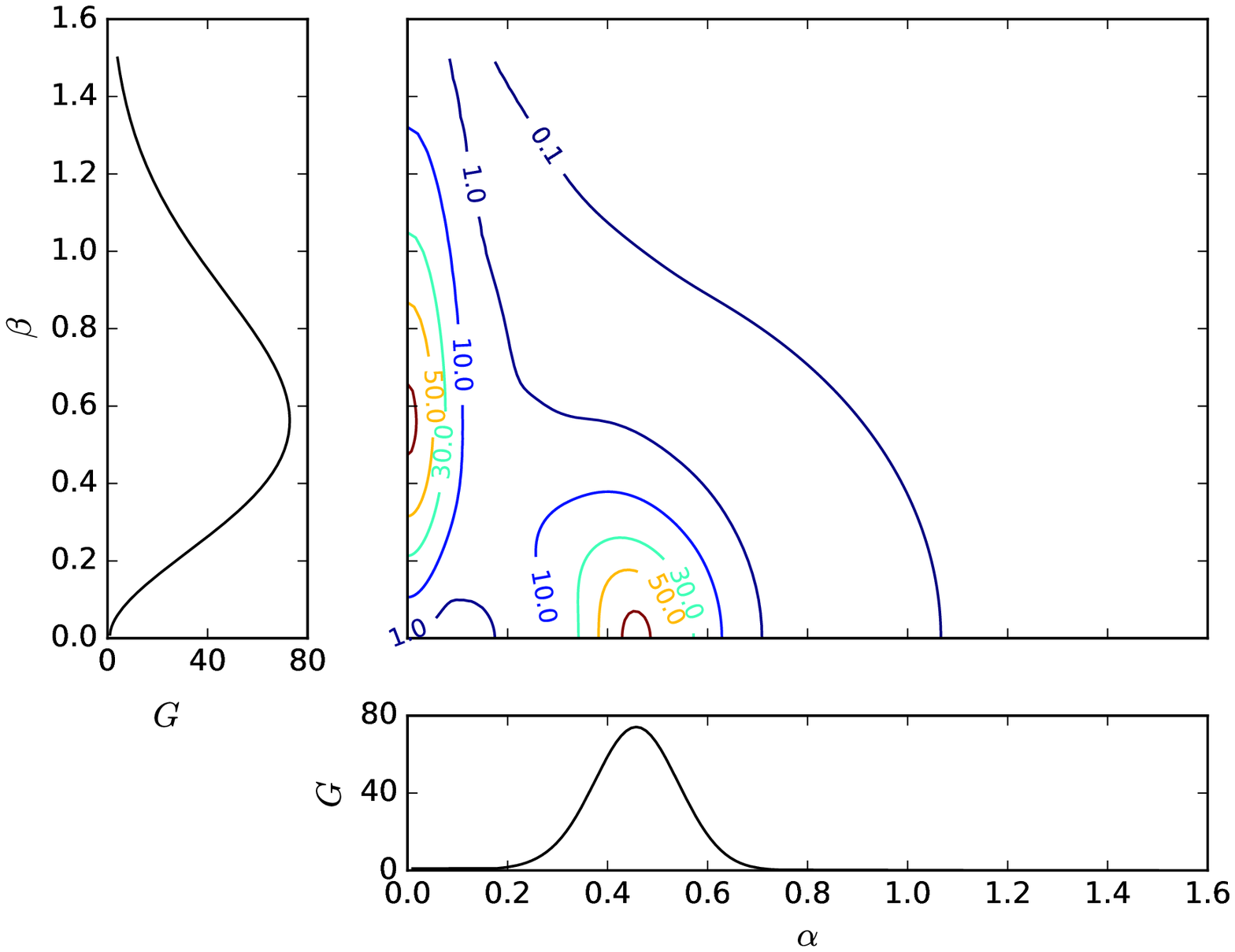}
    \caption{ $ 2t_0/\Rey = 0 $, $ 2t_1/\Rey = 2 $ }
    \label{fig:contourC}
  \end{subfigure} \nolinebreak
  \begin{subfigure}[b]{0.5\textwidth}
    \includegraphics[width=\textwidth]{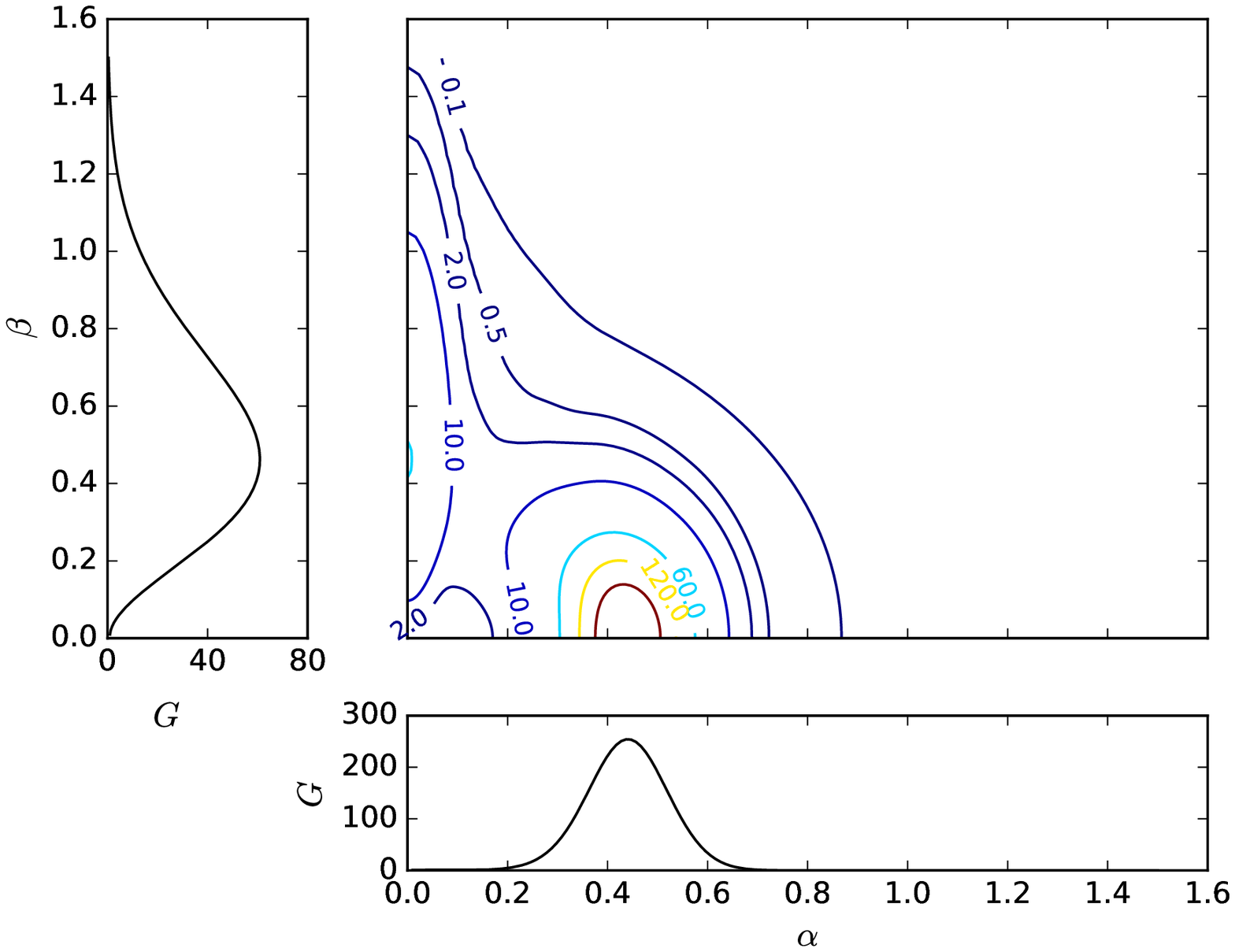}
    \caption{ $ 2t_0\Rey = 0 $, $ 2t_1/\Rey = 3 $ }
    \label{fig:contourD}
  \end{subfigure}
  \caption{Isolines of the amplification 
    $ G(\alpha,\beta,t_0,t_1 , \Rey_\delta = 316)$  for the boundary layer flow under a solitary wave, for different values of $ t_0 $ and $ t_1 $. The plots to the left and below the contour plot show a slice along the $ \beta $- and
$ \alpha $-axes, respectively.}
\label{fig:soliton2}
\end{figure}

A nonmodal stability analysis of this flow has been performed by \cite{VerschaevePedersenTropea2018} who found that for early times, streamwise streaks dominate, whereas for later times, during deceleration, nonmodal Tollmien-Schlichting waves dominate. In the following, we shall employ the present theoretical findings to investigate the case $ \Rey = 316 $, which, among other cases, has also been investigated in \cite{VerschaevePedersenTropea2018}. In figure \ref{fig:soliton2}, contour plots of the amplification $ G $ in Fourier space for different times $ t_0 $ and $ t_1 $ are plotted. During acceleration from $ 2t_0/\Rey = - 1 $ to $ 2t/\Rey = 0 $,
figure \ref{fig:soliton2}(a), the only optimal perturbations displaying growth are streamwise streaks with a maximum $ G \approx 15 $. For $ \beta = 0 $, no growth is visible. However, when integrating the system (\ref{eq:sys1}-\ref{eq:sys2}) for the same duration, but in the deceleration region from $ 2t_0/\Rey = 0 $ to $ 2t_1/\Rey = 1$, figure \ref{fig:soliton2}(b), we observe weak growth for optimal perturbations with $ \beta = 0 $ ($ G \approx 1.8 $). On the other hand, also streamwise streaks display larger amplifications than during acceleration, with a maximum at $ G \approx 80 $. However, when increasing the integration interval to $ 2t/\Rey = 2 $ and $ 2t/\Rey = 3 $, figures \ref{fig:soliton2}(c) and \ref{fig:soliton2}(d), respectively, streamwise streaks show slow decay from their peak at $ 2t/\Rey = 1 $, whereas two-dimensional perturbations display strong growth, centered around $ \alpha \approx 0.41 $.

\begin{figure} 
    \includegraphics[width=\textwidth]{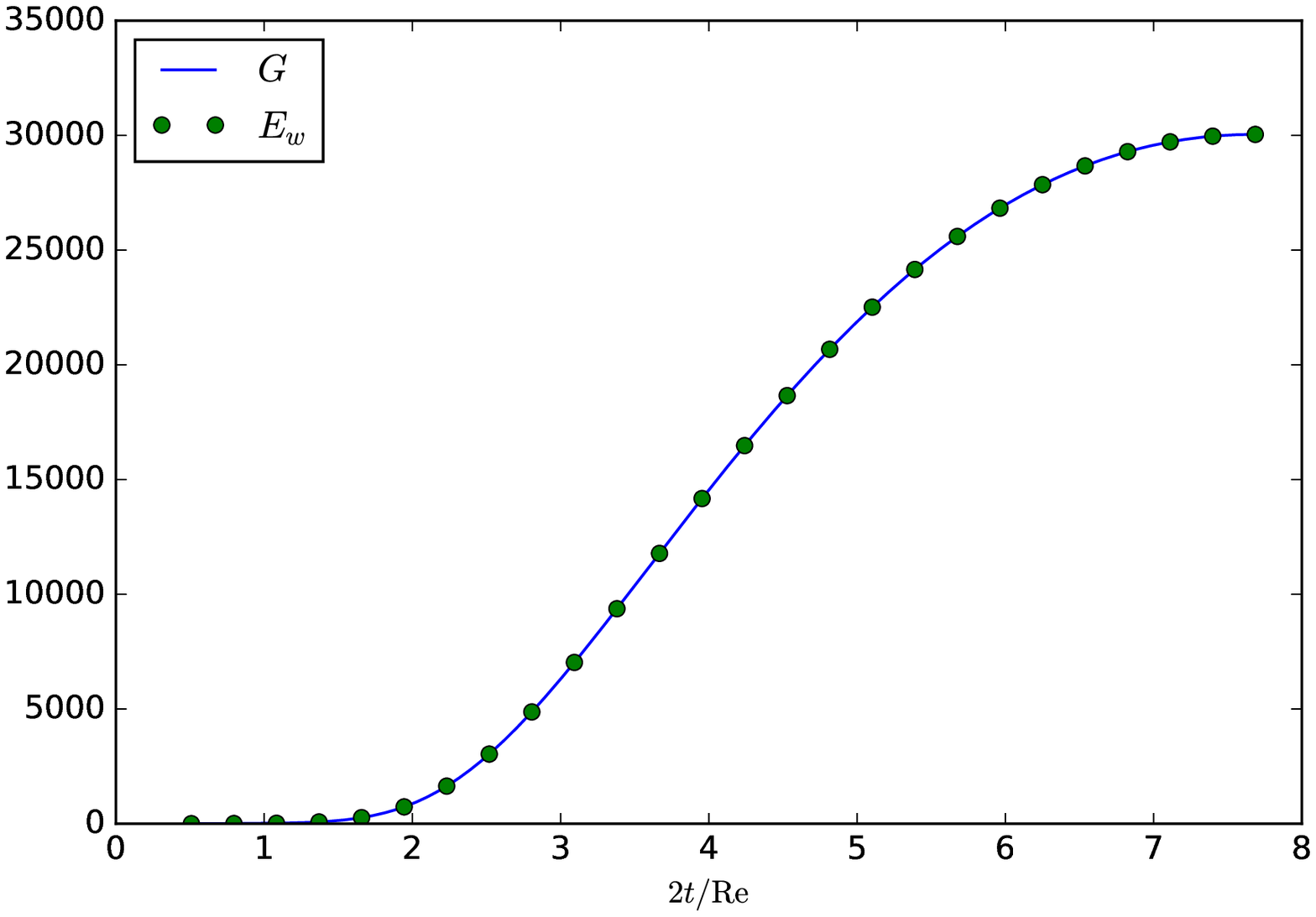}
    \caption{Amplification $ G $ of the optimal perturbation and temporal evolution of the amplification $ E_w $ of the perturbation leading to a maximum amplification at $ 2t/\Rey = 7.686 $ with $ 2t_0/\Rey = 0.509 $, $\alpha = 0.369$ and $ \beta = 0 $ for 
the boundary layer flow under a solitary wave at $ \Rey = 316 $.}
\label{fig:soliton3}
\end{figure}

\begin{figure} 
    \includegraphics[width=\textwidth]{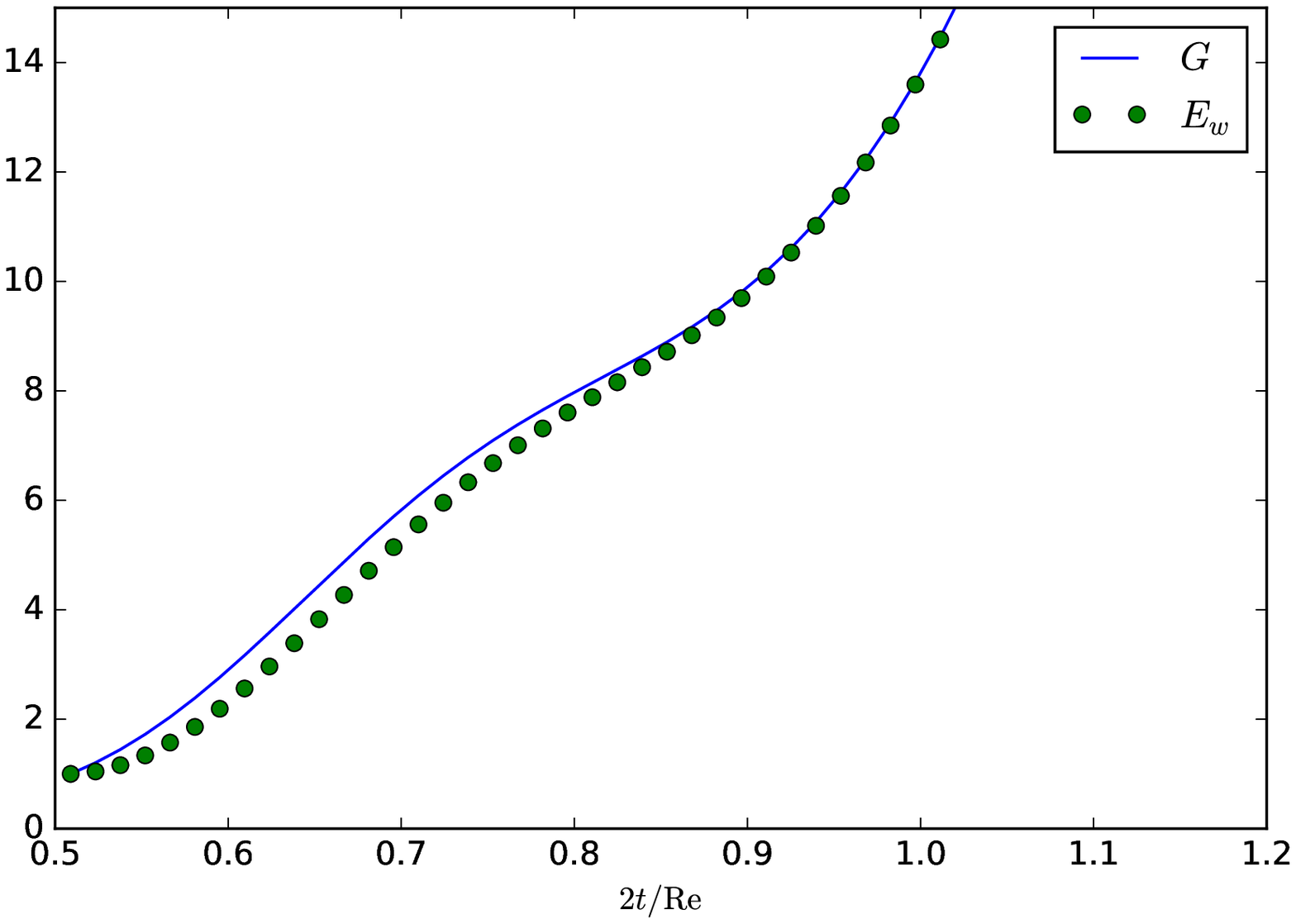}
    \caption{Zoom into figure \ref{fig:soliton3}. Amplification $ G $ of the optimal perturbation and temporal evolution of the amplification $ E_w $ of the perturbation leading to a maximum amplification at $ 2t/\Rey = 7.686 $ with $ 2t_0/\Rey = 0.509 $, $\alpha = 0.369$ and $ \beta = 0 $ for 
the boundary layer flow under a solitary wave at $ \Rey = 316 $.}
\label{fig:soliton4}
\end{figure}

\begin{figure} 
    \includegraphics[width=\textwidth]{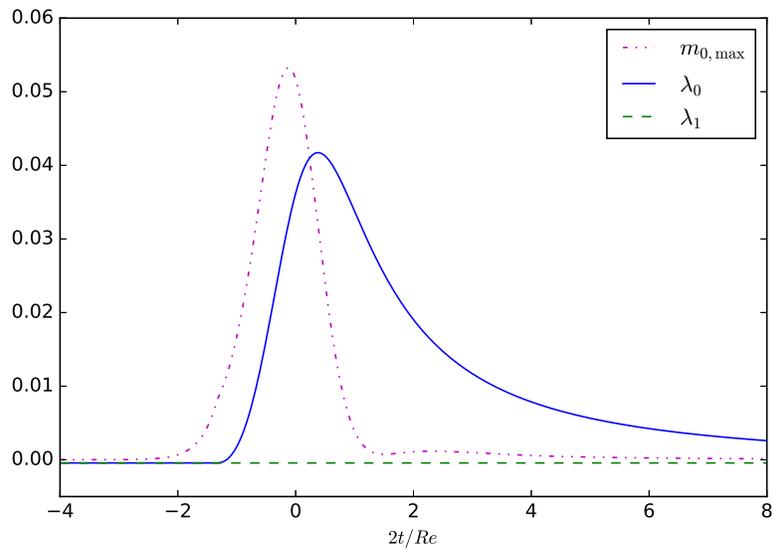}
    \caption{Temporal evolution of characteristic quantities of the DVK modes with $ \alpha = 0.369 $, $ \beta = 0 $, for the boundary layer flow under a solitary wave at $ \Rey = 316 $. Growth rates $ \lambda_0 $, $\lambda_1 $ and dispersion measure $ m_{0,\max} $. }
\label{fig:soliton5}
\end{figure}

\begin{figure} 
    \includegraphics[width=\textwidth]{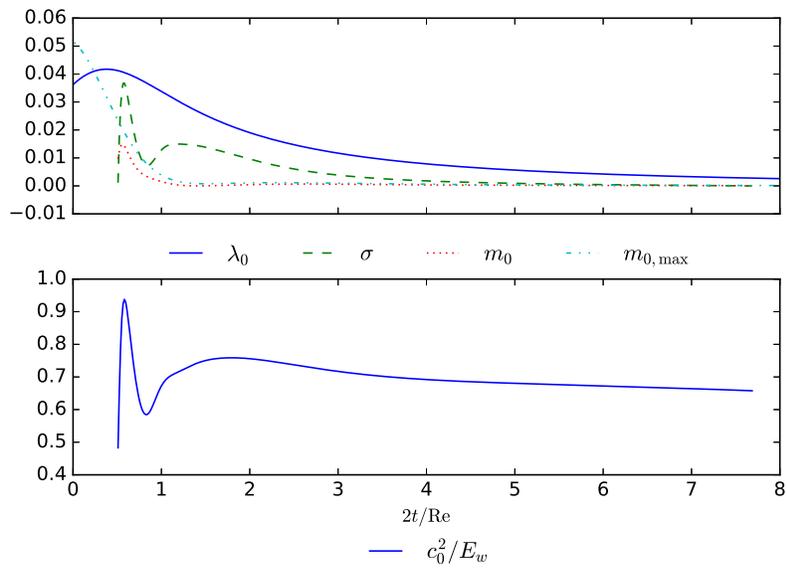}
    \caption{Temporal evolution of characteristic quantities of the nonmodal Tollmien-Schlichting wave with $ \alpha = 0.369 $, $ \beta = 0 $, for the boundary layer flow under a solitary wave at $ \Rey = 316 $. }
\label{fig:soliton6}
\end{figure}

\begin{figure} 
    \includegraphics[width=\textwidth]{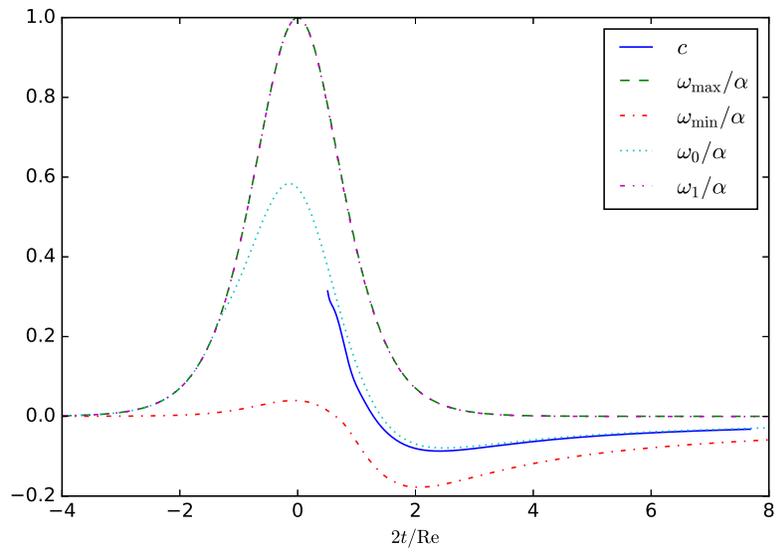}
    \caption{Boundary layer flow under a solitary wave with $ \Rey = 316 $. Displayed is the phase speed $ c $ of the nonmodal Tollmien-Schlichting wave with parameters given by equation (\ref{eq:phaseSpeed}) and phase speeds $ \omega_0/\alpha $ and $ \omega_1/\alpha $ of the first two DVK modes, equation (\ref{eq:baseFrequency}) and maximum and minimum phase speeds $ \omega_{\min}/\alpha $ and $ \omega_{\max}/\alpha $, equation (\ref{eq:boundOnFrequency}).}
\label{fig:soliton7}
\end{figure}

When maximizing the amplification of nonmodal Tollmien-Schlichting waves, we find
\be
\max_{\alpha,t_0,t_1} G( \alpha, \beta=0, t_0 , t_1, \Rey = 316)
= 3.0 \cdot 10^4,
\ee
where at maximum, we have
\bea
\alpha_{\max} &=& 0.369,  \label{eq:destruct1}\\
2t_{0,\max}/\Rey &=& 0.509, \\
2t_{1,\max}/\Rey &=& 7.686.  \label{eq:destruct2}
\eea
When plotting the amplification $ G $ of the optimal perturbation with $ \alpha = 0.369 $ and $ 2t_0/\Rey = 0.509 $ and the evolution of the energy $ E_w $ of the nonmodal Tollmien-Schlichting wave with parameters given by (\ref{eq:destruct1}-\ref{eq:destruct2}), cf. figure \ref{fig:soliton3}, we observe a qualitatively different picture than for Couette and Poiseuille flow, figures \ref{fig:Couette3} and \ref{fig:Poiseuille3}, respectively. The graphs of the temporal evolution of $ E_w $ and of $ G $ are lying on top of each other, except for a short initial period of time, cf. figure \ref{fig:soliton4} where a zoom of figure \ref{fig:soliton3} is displayed. This indicates that for most part of the deceleration region, the energy transfer optimization from and to the zeroth VDK mode as for Couette and Poiseuille is marginal. Instead, the nonmodal Tollmien-Schlichting wave evolves as if it had been an orthogonal mode with the largest growth rate. Figure \ref{fig:soliton5} lends some support to this behavior. We observe that for this Reynolds number, only the zeroth VKD mode displays regions of growth. As shall be discussed in appendix \ref{sec:continuous}, the first VKD mode, and so its eigenvalue $ \lambda_1 $, is probably already part of the continuous spectrum. The region of growth of $ \lambda_0 $ is skewed towards the deceleration region of the flow. In particular, we observe that growth starts at around $ 2t/\Rey \approx -1 $, but stretches much further into the deceleration region. However, probably more significant than the extended growth in the deceleration region is the skewness of the behavior of the dispersion measure $ m_{0,\max} $ which shows the opposite behavior. It displays much larger values in the acceleration region than in the deceleration region. In particular, it drops below the value of $ \lambda_0 $ at $ 2t /\Rey \approx 1/2 $ and stays at a low level for $ 2 t / \Rey > 1 $. This drop of dispersion of energy to and from higher VKD modes in the deceleration region of the flow, in combination with a significant growth rate, is probably at the origin of the behavior observed in figure \ref{fig:soliton3}. As a result, the mode $ \phi_0 $ behaves almost as an orthogonal mode evolving independently of the other VKD modes. However, this independence is not complete. In figure \ref{fig:soliton6}, several characteristic quantities are plotted for this nonmodal Tollmien-Schlichting wave. As to be expected, $ \sigma $ and $ m_0 $ honor their upper bounds $ \lambda_0 $ and $ m_{0,\max} $, respectively. On the other hand, the energy content for the zeroth VKD mode evolves around 70 \% of $ E_w $, indicating that, even in the case of reduced dispersion, the Tollmien-Schlichting wave transfers, along its course, some energy to higher VKD  which for the boundary layer flow under a solitary wave at $ \Rey = 316 $, are thought to lie in the continuous spectrum of equation (\ref{eq:davisPSE}). Whether this is due to a rest dispesion by the matrix $ \mathbf{N} $ or by the matrix $ \mathbf{F} $, which is small in magnitude but not zero in this case, remains an open question. Concerning the phase speed $ c $ of the nonmodal Tollmien-Schlichting wave, equation (\ref{eq:phaseSpeed}), its graph in figure \ref{fig:soliton7} shows that the nonmodal Tollmien-Schlichting wave travels approximately with the phase speed $ \omega_0/\alpha $ of the zeroth VKD mode, equation (\ref{eq:baseFrequency}). The upper and lower bounds for the phase speed, $ \omega_{\min} $ and $ \omega_{\max} $, respectively, cf. equation (\ref{eq:boundOnFrequency}), are computed by searching for the maximum and mininum eigenvalue of $ \mathbf{N} $. The upper bound seems to follow the speed of the outer flow, whereas the lower bound takes into account that the adverse pressure gradient causes a reverse flow, allowing the perturbations to travel in opposite direction. As the continuous spectrum typically attenuates inside the boundary layer, cf. reference \cite{HackZaki2012}, it is consistent that $ \phi_1 $ travels at the speed of the outer flow.

\section{Conclusions} \label{sec:conclusion}

In the present treatise, we showed that the parabolized stability equation approach derived in \cite{VerschaevePedersenTropea2018} leads to a Hermitian eigenvalue equation, equation (\ref{eq:davisPSE}) for the growth rate of the perturbation. The resulting set of orthonormal eigenfunctions, whose temporal continuation we called VKD modes, allowed us to formulate the governing equation in Heisenberg form. The resulting matrix equation consists of a Hermitian part responsible for growth of the perturbation and an skew-Hermitian part redistributing energy between VKD modes. Different quantities and bounds measuring the dispersion properties between VKD modes have been derived. The theoretical framework developed in the present treatise has been applied to three shear flows, Couette flow, Poiseuille flow and the boundary layer flow under a solitary wave. Two different regimes of a energy transfer between the VKD modes have been observed. For Couette and Poiseuille flow, with relatively large dispersion measure $ m_{0,\max}$, equation (\ref{eq:measure}), the optimal perturbation results from balancing transfer of energy from and to growing VKD modes in such a way that growth is largest. On the other hand, for the adverse pressure region of the boundary layer flow, a different regime of growth became visible. As the dispersion in this case is weak, growth is almost entirely provided by energy extraction from the base flow. As we have seen some of the energy accumulated is dispersed to the continuous spectrum. As a matter of fact, the present work leads to further questions, summarized in the following points:
\begin{itemize}
\item A underlying assumption of the present treatise is that growth and dispersion break even for some specific value of $ \alpha $ such that the resulting optimal perturbation reaches the global maximum for two-dimensional perturbations. Future research might find more specific conditions when and how this happens.
\item A better analytic bound for $ m_{0,\max}$ on the right hand side of equation (\ref{eq:measure}) might furnish us with more insight on how the shape of the base flow profile influences dispersion of energy to and from higher VKD modes.
\item Future research should deal with the improvement of the mathematical foundations of second and fourth order Sturm-Liouville problems with the eigenvalue term being a second order differential operator. In particular, semi-infinite intervals play a major role for boundary layer flows.
\item As the Heisenberg formulation, equation (\ref{eq:Heisenberg0}), allows to highlight the analogy to a quantum electrodynamical system, it might be worthwhile to consider techniques of this field to the present problem. In particular, it might be possible to model the dispersion effect of the higher VKD modes by a mean-field theory leading to an effective growth rate.
\end{itemize}
\vspace{0.5cm}
The author would like to thank Graigory Sutherland, Pawe{\l} Wroniszewski and Florian Schwertfirm for their support and encouragement.

\begin{appendix}

\section{Computing VKD-modes} \label{sec:modeContinuation}

The numerical computation of VKD-modes requires some care. In the following, we shall elucidate two important issues when dealing with VKD-modes.

\subsection{Normalizing VKD modes in time}

The eigenfunctions resulting equation (\ref{eq:davisPSE}) are only determined up to a multiplicative constant. In order to find the correct scaling of the eigenfunctions in time, we return to constraint (\ref{eq:constraint}). By means of a semi-discrete version of constraint (\ref{eq:constraint}) at the midpoint $ (t_1 + t_2)/2 $, the mode $ \phi $ at time $ t_2 $ is normalized using its value at $ t_1 $:
\bea
\lefteqn{ \frac{1}{2(t_2-t_1)} \int \limits_a^b \left( D \phi^\dagger(t_2) + D\phi^\dagger(t_1) \right) \left( D \phi(t_2) - D \phi(t_1)  \right) } \nonumber \\ 
& & \quad \quad \quad \quad \quad \quad \quad \quad \quad+ k^2 \left( \phi^\dagger(t_2) + \phi^\dagger(t_1) \right) \left( \phi(t_2) -  \phi(t_1)  \right) \, dz = 0
\label{eq:constraintDiscrete}
\eea
The real and imaginary part of equation (\ref{eq:constraintDiscrete}) are given by:
\bea
\int \limits_a^b \left( | D\phi(t_2)|^2 + k^2 | \phi(t_2) |^2 \right)
- \left( | D\phi(t_1)|^2 + k^2 | \phi(t_1) |^2 \right) \, dz &=& 0 \label{eq:discrete1} \\
\int \limits_a^b \left( D\phi^\dagger(t_1) D\phi(t_2) + k^2 \phi^\dagger(t_1)\phi(t_2) \right) - \left( D\phi^\dagger(t_2) D\phi(t_1) + k^2 \phi^\dagger(t_2)\phi(t_1) \right) \, dz &=& 0\label{eq:discrete2} 
\eea
Equation (\ref{eq:discrete1}) represents the conservation of energy, whereas equation (\ref{eq:discrete2}) can be written as
\be
A - A^\dagger = 0,
\ee 
implying that the imaginary part of $ A $ vanishes. We can thus write:
\bea
A &=& A_r + {\rm i } A_i \\
&=&  \int \limits_a^b \left( D\phi^\dagger(t_1) D\phi(t_2) + k^2 \phi^\dagger(t_1)\phi(t_2) \right) \, dz \\
&=& \int \limits_a^b  D\phi_r(t_1) D \phi_r(t_2) + D\phi_i(t_1)D\phi_r(t_2) + k^2 \left( \phi_r(t_1) \phi_r(t_2) + \phi_i(t_1)\phi_i(t_2) \right) \, dz \nonumber \\
& & + {\rm i} \int \limits_a^b  D\phi_r(t_1) D \phi_i(t_2) - D\phi_i(t_1)D\phi_r(t_2) + k^2 \left( \phi_r(t_1) \phi_i(t_2) - \phi_i(t_1)\phi_r(t_2) \right) \, dz 
\eea
After having traced the eigenfunction $ \tilde{\phi} $ at $ t_2 $ corresponding to the mode at $ t_1 $, we can pose:
\be
 \phi(t_2) = e^{ {\rm i} \delta } \tilde{\phi}(t_2) ,
\ee
where we have assumed that $ \tilde{\phi}(t_2) $ is already normalized by the energy. We are thus left determining the phase $ \delta $ such that (\ref{eq:discrete2}) is satisfied. This is obtained for
\be
\tan \delta = - \frac{ \tilde{A}_i}{\tilde{A}_r},
\ee
where $ \tilde{A}_r $ and $ \tilde{A}_i $ correspond to $ A_r $ and $ A_i $ with only $ \phi(t_1) $ replaced by $ \tilde{\phi}(t_1) $.

\subsection{Infinite domains} \label{sec:continuous}

\begin{figure} 
  \centering
  \begin{subfigure}[b]{0.5\textwidth}
    \includegraphics[width=\textwidth]{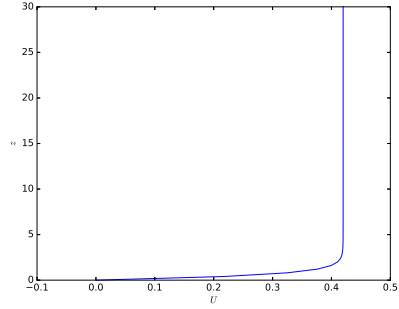}
    \caption{ $ 2t/\Rey = -1 $}
  \end{subfigure} \nolinebreak
  \begin{subfigure}[b]{0.5\textwidth}
    \includegraphics[width=\textwidth]{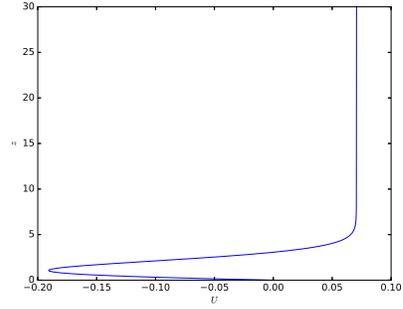}
    \caption{ $ 2t/\Rey = 2 $}
  \end{subfigure}
  \caption{Velocity profile of the boundary layer under a solitary wave at $ 2t/\Rey = -1 $ and $ 2t/\Rey = 2 $ }
  \label{fig:DVK-modes2}
\end{figure}

\begin{figure} 
  \centering
  \begin{subfigure}[b]{0.5\textwidth}
    \includegraphics[width=\textwidth]{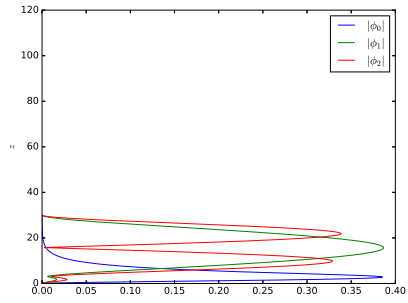}
    \caption{ $ h = 30 $}
  \end{subfigure} \nolinebreak
  \begin{subfigure}[b]{0.5\textwidth}
    \includegraphics[width=\textwidth]{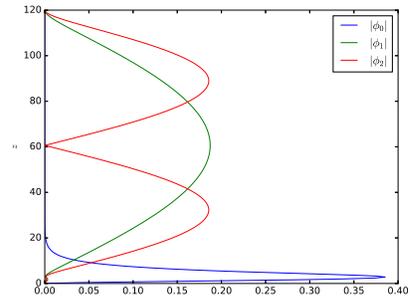}
    \caption{ $ h = 120 $}
  \end{subfigure} 
  \caption{Magnitude of the first three DVK modes with $ \alpha = 0.369 $ and $ \beta = 0$ for the boundary layer flow under a solitary wave for $ \Rey = 316 $ at $ 2t/\Rey = 2 $. The numerical domain is truncated at different values of $ h $.}
\label{fig:DVK-modes3}
\end{figure}

\begin{figure} 
  \centering
  \begin{subfigure}[b]{0.5\textwidth}
    \includegraphics[width=\textwidth]{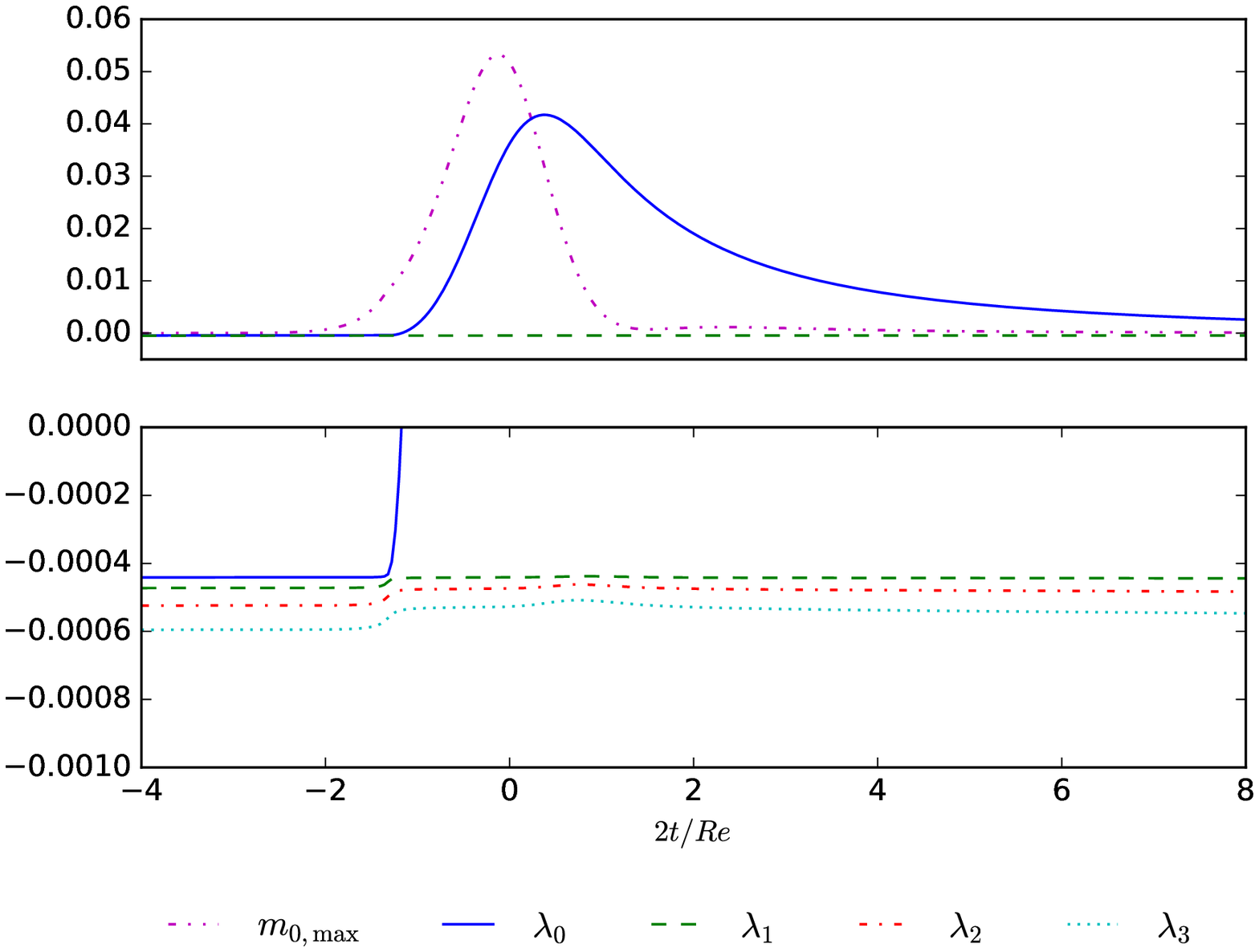}
    \caption{ $ h = 30 $}
  \end{subfigure} \nolinebreak
  \begin{subfigure}[b]{0.5\textwidth}
    \includegraphics[width=\textwidth]{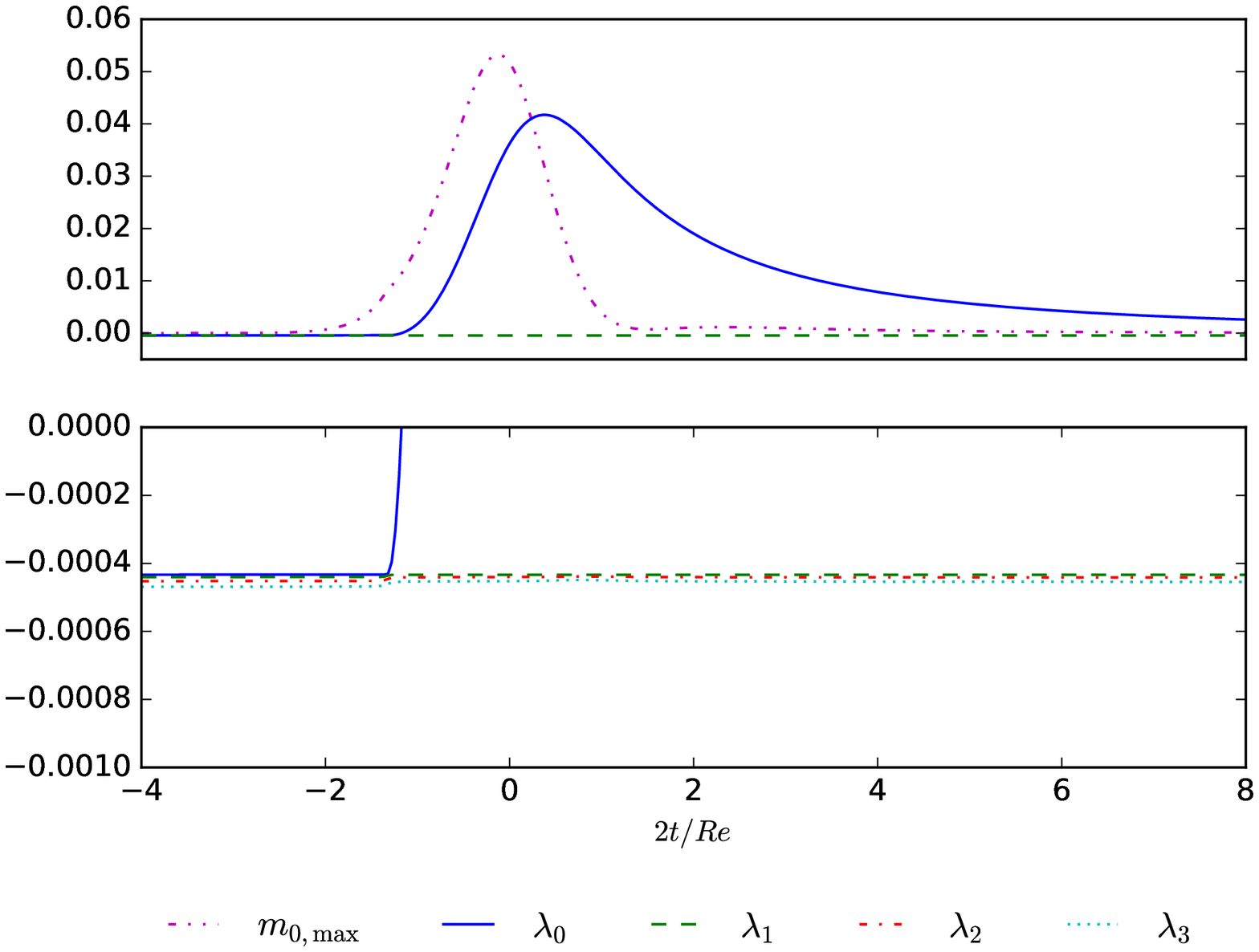}
    \caption{ $ h = 120 $}
  \end{subfigure} 
  \caption{Temporal evolution of characteristic quantities of the nonmodal Tollmien-Schlichting wave with $ \alpha = 0.369 $, $ \beta = 0 $, for the boundary layer flow under a solitary wave at $ \Rey = 316 $. The numerical domain is truncated at different values of $ h $. }
\label{fig:DVK-modes5}
\end{figure}

\begin{figure} 
  \centering
  \begin{subfigure}[b]{0.5\textwidth}
    \includegraphics[width=\textwidth]{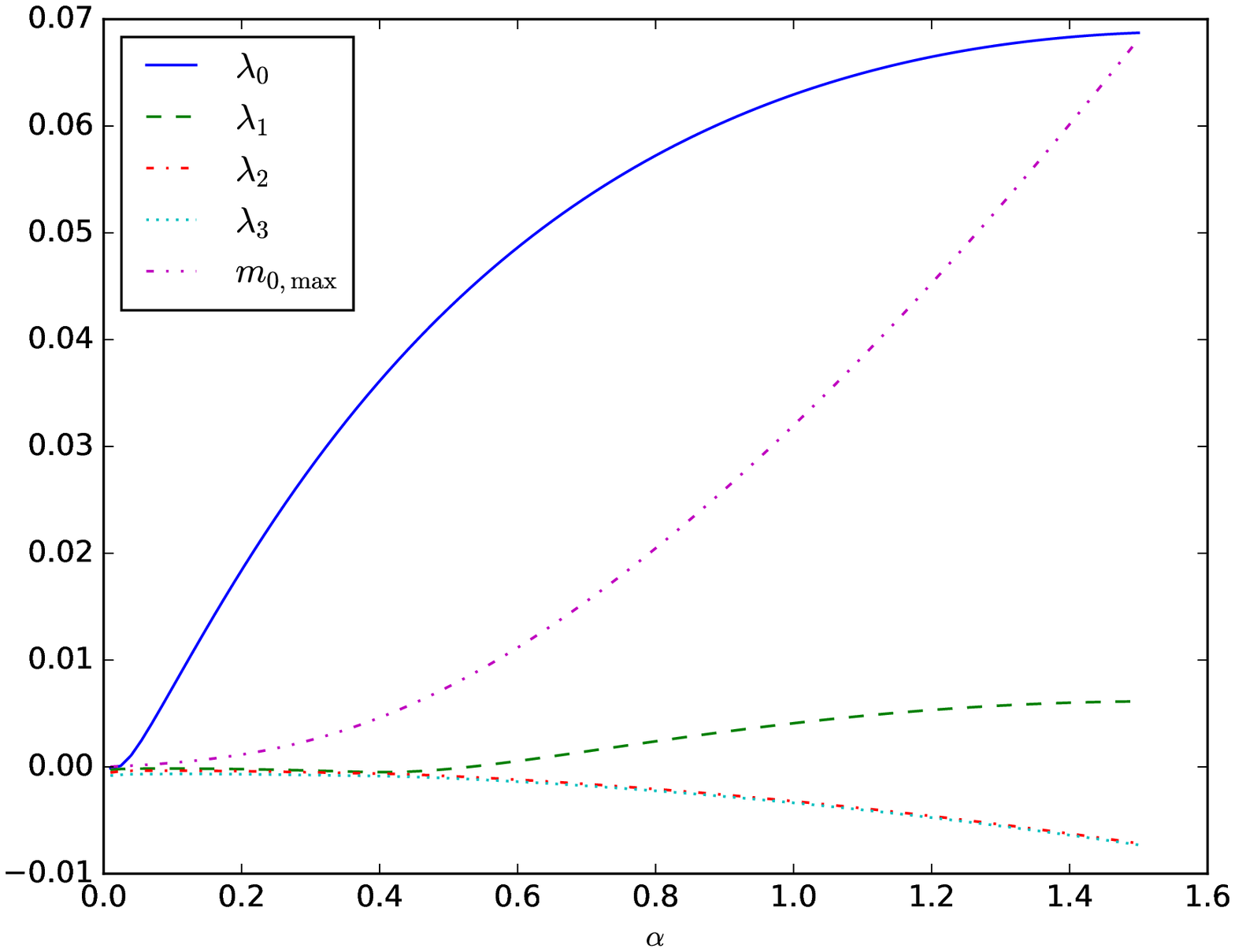}
    \caption{ $ h = 30 $}
    \label{fig:lambdaVersusAlpha_1}
  \end{subfigure} \nolinebreak
  \begin{subfigure}[b]{0.5\textwidth}
    \includegraphics[width=\textwidth]{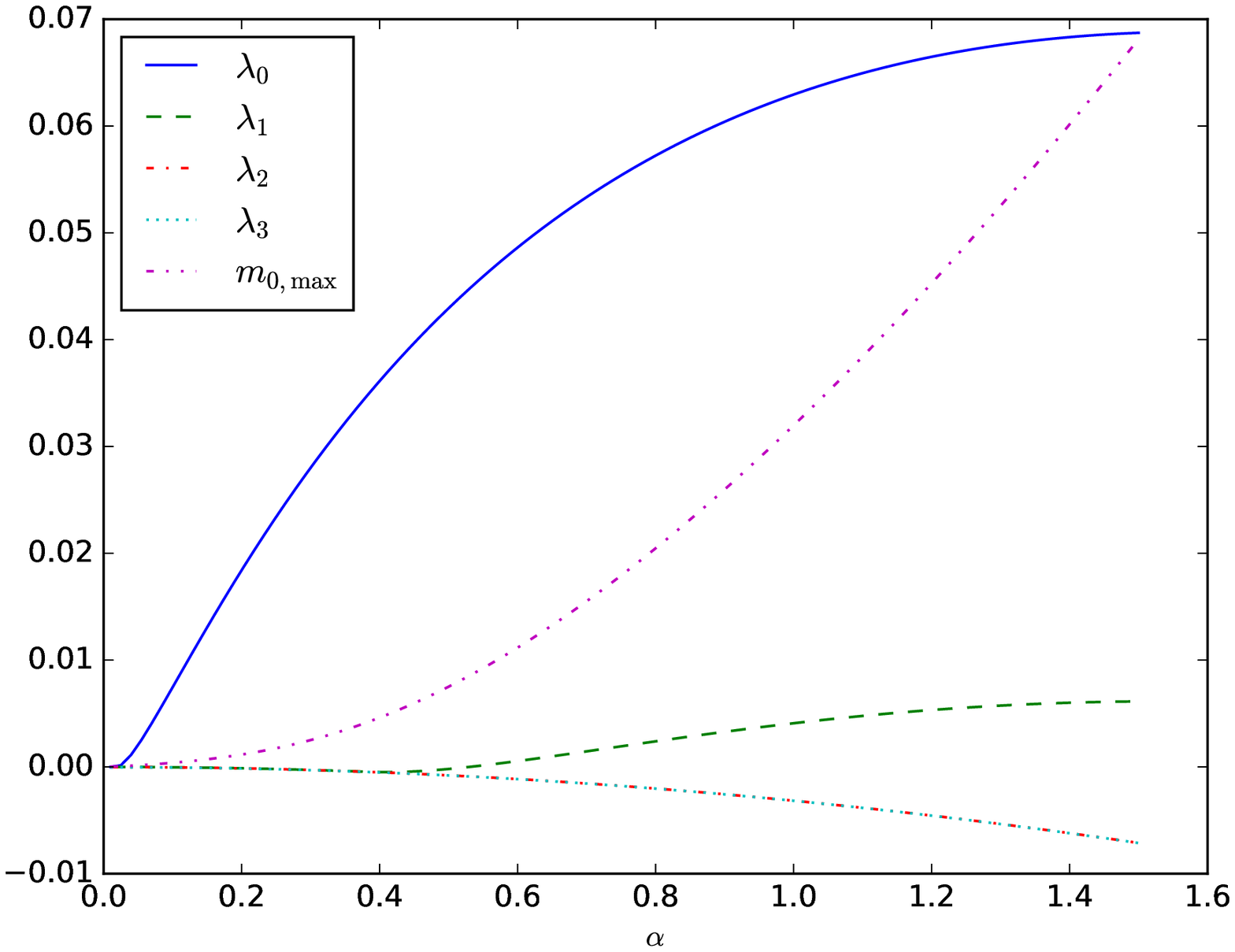}
    \caption{ $ h = 120 $}
    \label{fig:lambdaVersusAlpha1}
  \end{subfigure} 
  \caption{Growth rates $ \lambda_i $ and dispersion measure
    $ m_{0,\max }$ for the boundary layer flow under a solitary wave in function 
    of $ \alpha $ at time  $ 2 t / \Rey = 1 $. 
The numerical domain is truncated at different values of $ h $.}
\label{fig:soliton8}
\end{figure}

\begin{figure}
\centering
\includegraphics[width=\textwidth]{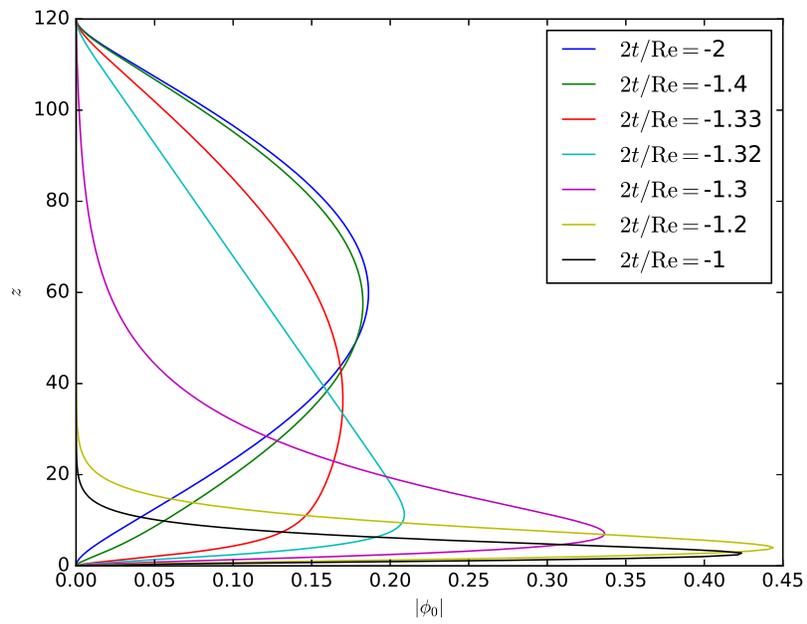}
\caption{Magnitude of the zeroth DVK mode for the boundary layer flow under a solitary wave for $ \Rey = 316 $ for different values of time.}
\label{fig:DVK-modes4}
\end{figure}

As mentioned in section (\ref{sec:results}), when considering flows in infinite or semi-infinite intervals, we are not in possession of any theoretical result giving an estimate on the number of discrete eigenvalues. Therefore, we need to carefully check the numerical results when solving equation (\ref{eq:davisPSE}) by varying the number of basis polynomials and the extend $ h $ at which we truncate the numerical domain. In the following, we consider VKD modes for the boundary layer flow under a solitary wave at two points in time, $ 2t/\Rey = -1 $ and $ 2t/\Rey = 2 $. From figure \ref{fig:DVK-modes2}, we infer that the boundary layer itself has a thickness of around $ 3 $ and $ 6 $ respectively. In figure \ref{fig:DVK-modes3}, the magnitude of the first three VKD modes is plotted with respect to the wall normal distance for two different values of the numerical cut-off parameter $ h $. The solutions are well converged concerning the number of polynomials used, ie. 129.The spatial extend of the zeroth VKD-mode seems to be well resolved for both values of $ h $, as it falls off exponentially to zero in $ z $. However, the first and second VKD mode do not display such a drop off but fill all the domain given to them, also when doubling $ h $ once more (figure not shown). This suggests that for $ \Rey = 316 $, $ \alpha = 0.369 $ and $ 2t/\Rey = 2$, equation (\ref{eq:davisPSE}) only possesses a single discrete eigenvalue and eigenfunction. All other eigenvalues and eigenfunctions obtained have to be considered numerical artifacts. This can also be observed for the eigenvalues, cf. figure \ref{fig:DVK-modes5}, where only $ \lambda_0 $ is well converged with respect to $ h $, whereas the others gather closer together for increasing $ h$. However, when increasing $ \alpha $, figure \ref{fig:soliton8}, we observe that $ \lambda_1 $ is bifurcating from the bundled eigenvalues and becoming positive, indicating the existence of a second discrete mode for $ \Rey = 316 $ and $ 2t/\Rey = 1$. As a matter of fact $ \phi_1 $ and $ \phi_2 $ in figure \ref{fig:DVK-modes2} resemble more functions typical for the continuous spectrum, cf. reference \cite{HackZaki2012}, displaying oscillations in the free stream and attenuating inside the boundary layer. For the flow in question, the zeroth VKD-mode, $ \phi_0 $, obtains its characteristic shape at around $ 2t/\Rey = - 1.3 $, cf. figure \ref{fig:DVK-modes4}. As can be observed from figure \ref{fig:DVK-modes4}, for 
$ 2t/\Rey = -1 $, the eigenfunction $ \phi_0 $ displays a single antinode over the entire domain similar to $ \phi_1 $ at $ 2t/\Rey = 2 $ in figure \ref{fig:DVK-modes2}. Gradually, this antinode morphs into the exponential hump with an extend comparable to the boundary layer thickness. Similarly, all other eigenfunctions lose one of their antinode at this point in time (not shown).

\section{Path integral expansion}  \label{sec:pathIntegral}

\begin{figure} 
\centering
\includegraphics[width=\textwidth]{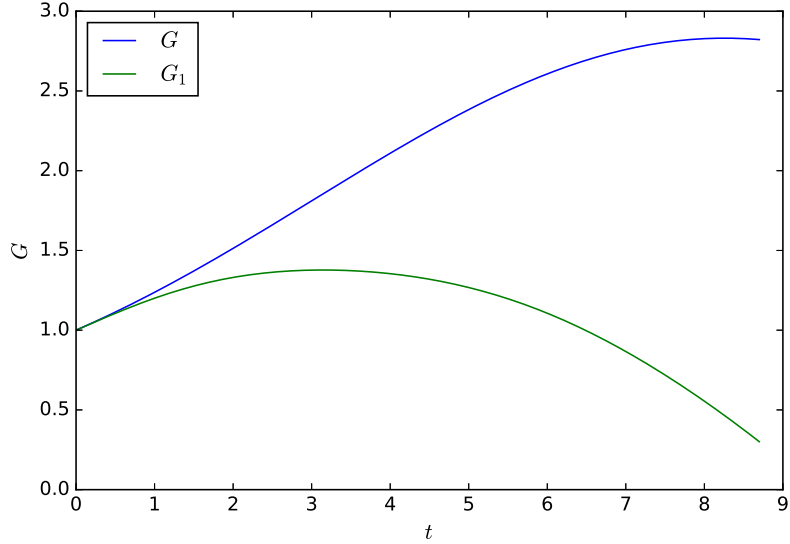}
\caption{Amplification $ G $ of the optimal perturbation and approximation $ G_1 $, equation (\ref{eq:pathIntegralG1}) for $ \alpha = 1.21 $ and $ \beta = 0 $ for Couette flow at $ \Rey = 1000 $.}
\label{fig:pathIntegral}
\end{figure}

As touched upon in section \ref{sec:results}, equation (\ref{eq:Heisenberg}) modeling the nonmodal Tollmien-Schlichting waves as a linear combination of VKD modes, can be considered a system composed of a base Hamiltonian and a perturbation potential. In the following, we shall assume for simplicity that the base flow is steady and thus matrices $ \mathbf{\Lambda} $ and $ \mathbf{N} $ are constant in time. Matrix $ N $ can be diagonalized:
\be
\mathbf{N} = \mathbf{S} \mathbf{H} \mathbf{S}^T,
\label{eq:diagonalization}
\ee
where $ \mathbf{H} $ is a diagonal matrix with real eigenvalues on its diagonal. This allows us to write equation (\ref{eq:Heisenberg}) as
\be
{\mathbf{\dot{d}}} = {\rm i} \mathbf{H} \mathbf{d} + \mathbf{V} \mathbf{d},
\label{eq:HeisenbergAppendix1}
\ee
where we have written:
\bea
\mathbf{d} &=& \mathbf{S}^T \mathbf{c} \\
\mathbf{V} &=& \mathbf{S}^T \mathbf{\Lambda} \mathbf{S}
\eea
Likewise we have for the fundamental solution $ \mathbf{X} $:
\be
{\mathbf{\dot{X}}} = {\rm i} \mathbf{H} \mathbf{X} + \mathbf{V} \mathbf{X},
\label{eq:HeisenbergAppendix2}
\ee
An integrating factor for equation (\ref{eq:HeisenbergAppendix2}) is given by
\be
\mathbf{X} = e^{ {\rm i} \mathbf{H} t } \mathbf{Y},
\ee
allowing us to write equation (\ref{eq:HeisenbergAppendix2}) as
\be
{\mathbf{\dot{Y}}} = \underbrace{e^{ -{\rm i} \mathbf{H} t } \mathbf{V} e^{ {\rm i} \mathbf{H} t }  }_{= \mathbf{U}(t) } \mathbf{Y},
\label{eq:HeisenbergAppendix3}
\ee
Path integral formulations, cf. reference \cite{MartinRothen1990}, allow us to find approximations for $\mathbf{Y}$. Integrating equation (\ref{eq:HeisenbergAppendix2}) from $ t_0 $ to $ t $, we obtain:
\be
\mathbf{Y}(t,t_0) = \mathbf{I} + \int \limits_{t_0}^t  \, dt_1 \mathbf{U}(t_1) \mathbf{Y}(t_1,t_0)
\label{eq:HeisenbergAppendix4}
\ee
Substituting $ \mathbf{Y} $ back into (\ref{eq:HeisenbergAppendix3}) and performing repeated integrations leads to an expansion formula for $ \mathbf{Y}$:
\bea
\mathbf{Y} &=& \mathbf{I} + \underbrace{\int \limits_{t_0}^t  \, dt_1 \mathbf{U}(t_1)}_{\mathbf{A}} +
\underbrace{\int \limits_{t_0}^t  \, dt_1 \int \limits_{t_0}^{t_1}  \, dt_2 \mathbf{U}(t_1)\mathbf{U}(t_2)}_{\mathbf{B}} + \ldots \label{eq:HeisenbergAppendix5}\\
&=&  \sum \limits_{n=0}^\infty \int \limits_{t_0}^{t} \, dt_1 \ldots \int \limits_{t_0}^{t_{n-1}} dt_{n} \, \mathbf{U}(t_1) \ldots \mathbf{U}(t_{n}). \label{eq:HeisenbergAppendix6}
\eea
The elements of $ \mathbf{A} $ and $ \mathbf{B} $ are given explicitly as:
\bea
A_{nm}  &=& \left( \int \limits_{t_0}^t \, dt_1 \mathbf{U}(t_1) \right)_{nm}\\
& = & \left\{ \begin{array}{cc} V_{nn} (t - t_0 ) & n = m \\
  \frac{1}{{\rm i}\left( h_m - h_n \right) } V_{nm}
  \left[ e^{ {\rm i} \left( h_m - h_n \right) t } - e^{ {\rm i} \left( h_m - h_n \right) t_0 }\right] & n \neq m
  \end{array} \right. \\
B_{nm} &= & \left( \int \limits_{t_0}^t \, dt_1 \int \limits_{t_0}^{t_1} \, dt_2 \mathbf{U}(t_1) \mathbf{U}(t_2) \right)_{nm} \label{eq:pairwise} \\
&=& \left\{ \begin{array}{cc} \begin{array}{c}
    \sum \limits_{k\neq n} \frac{ |V_{nk} |^2}{ (h_n - h_k )^2}
    \left[ 1 + {\rm i} \left(  h_k - h_n \right)  ( t - t_0 ) 
      - e^{ {\rm i} ( h_k - h_n ) ( t-t_0) } \right] \\
    + \frac{1}{2} | V_{nn} |^2 \left( t - t_0 \right)^2 \end{array} & n = m \\
  \begin{array}{c} \sum \limits_{k\neq n,m}
    \frac{ V_{nk} V_{km} } { ( h_m - h_k ) ( h_m - h_n ) (h_k - h_n )}
    \Big\{ (h_k -h_m ) e^{ {\rm i } ( h_m - h_n) t_0 } \\
    + ( h_m - h_n ) e^{ {\rm i } (h_k - h_n ) t +
      { \rm i} ( h_m - h_k ) t_0 } 
    + (h_n - h_k) e^{ {\rm i} ( h_m - h_n ) t }\Big\} \\
   + \frac{V_{nn} V_{nm}}{ ( h_m - h_n )^2} \Big\{ {\rm i} ( h_m - h_n ) ( t - t_0 ) e^{ {\rm i} ( h_m - h_n ) t_0 } \\
   - e^{ { \rm i} ( h_m - h_n ) t } + e^{ { \rm i} ( h_m - h_n ) t_0 } \Big\} \\
   + \frac{V_{nm} V_{mm}}{ ( h_m - h_n )^2} \Big\{ -{\rm i} (h_m - h_n) (t - t_0 ) e^{ {\rm i} ( h_m - h_n ) t} \\
+   e^{ {\rm i} ( h_m - h_n ) t } - e^{ {\rm i} ( h_m - h_n ) t_0 } \Big\}
  \end{array} & n \neq m \end{array} \right.  \label{eq:triple}
\eea
In quantum electrodynamics, the terms in formula (\ref{eq:HeisenbergAppendix5}) represent different levels of interaction between the modes of the base Hamiltonian $ \mathbf{H} $, equation (\ref{eq:diagonalization}). The first term on the right hand side of
(\ref{eq:HeisenbergAppendix5}) represents the undisturbed solution to the base Hamiltonian. The matrix $ \mathbf{A} $, equation (\ref{eq:pairwise}), on the other hand, stands for the effect of pairwise interactions between modes under action of the potential $ \mathbf{V} $. Likewise, the matrix $ \mathbf{B} $, equation (\ref{eq:triple}), accounts for triple interactions between modes under action of the potential $ \mathbf{V} $, and so on. The energy $ \mathbf{c}^\dagger \mathbf{c} $ of the nonmodal Tollmien-Schlichting wave is given by:
\bea
\mathbf{c}^\dagger \mathbf{c} &=& \mathbf{d}^\dagger \mathbf{S}^\dagger \mathbf{S} \mathbf{d} \\
&=& \mathbf{d}^\dagger \mathbf{d} \\
&=& \mathbf{d}^\dagger_0 \mathbf{X}^\dagger \mathbf{X} \mathbf{d}_0 \\
&=& \mathbf{d}^\dagger_0 \mathbf{Y}^\dagger \mathbf{Y} \mathbf{d}_0 \\
&=& \mathbf{d}^\dagger_0 \left( \mathbf{I} +  \mathbf{A}^\dagger + \mathbf{B}^\dagger + \ldots \right) \left( \mathbf{I} +  \mathbf{A} + \mathbf{B} + \ldots \right) \mathbf{d}_0 \\
&=& \mathbf{d}^\dagger_0 \mathbf{d}_0 + 
\mathbf{d}^\dagger_0  \left( \mathbf{A}^\dagger + \mathbf{A} \right)\mathbf{d}_0 \nonumber \\
& & \quad +
\mathbf{d}^\dagger_0  \left( \mathbf{A}^\dagger \mathbf{A} \right)\mathbf{d}_0 +
\mathbf{d}^\dagger_0  \left( \mathbf{B}^\dagger + \mathbf{B} \right)\mathbf{d}_0 +
\mathbf{d}^\dagger_0  \left( \mathbf{A}^\dagger \mathbf{B} + \mathbf{B}^\dagger \mathbf{A}\right)\mathbf{d}_0 + \ldots
\eea
By means of this formula, we can define successive approximations to $ G $:
\bea
G_0 & =& 1 \\
G_1 & =& \max_{\mathbf{d}_0} \frac{ \mathbf{d}_0^\dagger \left( I + \mathbf{A}^\dagger + \mathbf{A} \right)\mathbf{d}_0 } { \mathbf{d}_0^\dagger \mathbf{d}_0 } \label{eq:pathIntegralG1}\\
G_2 & =& \max_{\mathbf{d}_0} \frac{ \mathbf{d}_0^\dagger \left( I + \mathbf{A}^\dagger + \mathbf{A} + \mathbf{A}^\dagger \mathbf{A} + \mathbf{B}^\dagger + \mathbf{B} \right)\mathbf{d}_0 } { \mathbf{d}_0^\dagger \mathbf{d}_0 } \\
\ldots
\eea
In figure \ref{fig:pathIntegral}, the amplification $ G $ of the optimal perturbation is displayed next to its approximation $ G_1 $ for $ \alpha = 1.21 $ and $ \beta = 0 $ for Couette flow at $ \Rey = 1000 $. For small times the approximation $ G_1 $ follows the solution $ G $ closely, but then diverges for larger times. A typical issue for path integral approximations is that for higher orders the approximation diverges. This is also the case for nonmodal Tollmien-Schlichting waves, where $ G_2 $ diverges even for smaller times than $ G_1 $ due to the term $ \mathbf{A}^\dagger \mathbf{A} $. In fact, because of the viscous term, the matrix $ \mathbf{A} $ has large negative eigenvalues which leads to $ \mathbf{A}^\dagger \mathbf{A} $ having large positive eigenvalues causing unphysically large growth.

\end{appendix}


\begin{thebibliography}{10}

\bibitem{BertolottiHerbertSpalart1992}
F.~Bertolotti, T.~Herbert, and P.~Spalart.
\newblock Linear and nonlinear stability of the {B}lasius boundary layer.
\newblock {\em Journal of Fluid Mechanics}, 242:441--474, 1992.

\bibitem{BrandtSchlatterHenningson2004}
L.~Brandt, P.~Schlatter, and D.~S. Henningson.
\newblock Transition in boundary layers subject to free-stream turbulence.
\newblock {\em Journal of Fluid Mechanics}, 517:167--198, 2004.

\bibitem{ButlerFarrell1992}
K.~M. Butler and B.~F. Farrell.
\newblock Three-dimensional optimal perturbations in viscous shear flow.
\newblock {\em Physics of Fluids A}, 4:1637--1650, 1992.

\bibitem{Charru2011}
F.~Charru.
\newblock {\em Hydrodynamic instabilities}.
\newblock Cambridge University Press, 2011.

\bibitem{Chatterji1998}
S.~D. Chatterji.
\newblock {\em Cours d'Analyse 3: {E}quations diff{\'e}rentielles ordinaires et
  aux d{\'e}riv{\'e}es partielles}.
\newblock Presses polytechniques et universitaires romandes, 1998.

\bibitem{DavisKerczek1973}
S.~H. Davis and C.~von Kerczek.
\newblock A reformulation of energy stability theory.
\newblock {\em Archive for Rational Mechanics and Analysis}, 52(2):112--117,
  June 1973.

\bibitem{DrazinReid1981}
P.~G. Drazin and W.~H. Reid.
\newblock {\em Hydrodynamic Stability}.
\newblock Cambridge University Press, 1981.

\bibitem{Gustavsson1991}
H.~Gustavsson.
\newblock Energy growth of three-dimensional disturbances in plane {P}oiseuille
  flow.
\newblock {\em Journal of Fluid Mechanics}, 224:241--260, 1991.

\bibitem{HackZaki2012}
M.~J.~P. Hack and T.~A. Zaki.
\newblock The continuous spectrum of time-harmonic shear layers.
\newblock {\em Physics of Fluids}, 24:034101--1--23, 2012.

\bibitem{Jimenez2013}
J.~Jimenez.
\newblock How linear is wall-bounded turbulence?
\newblock {\em Physics of Fluids}, 25:110814--1--19, 2013.

\bibitem{LuchiniBottaro2014}
P.~Luchini and A.~Bottaro.
\newblock Adjoint equations in stability analysis.
\newblock {\em Annual Review of Fluid Mechanics}, 46:493--517, 2014.

\bibitem{MartinRothen1990}
P.~A. Martin and F.~Rothen.
\newblock {\em Probl{\`e}mes {\`a} {N}-corps et champs quantiques}.
\newblock Presses polytechniques et universitaires romandes, 1990.

\bibitem{Schmid2007}
P.~J. Schmid.
\newblock Nonmodal stability theory.
\newblock {\em Annual Review of Fluid Mechanics}, 39:129--162, 2007.

\bibitem{SchmidHenningson2001}
P.~J. Schmid and D.~S. Henningson.
\newblock {\em Stability and Transition in Shear Flows}.
\newblock Springer Science+Business Media, 2001.

\bibitem{Shen1994}
J.~Shen.
\newblock Efficient spectral-galerkin method i. direct solvers for the second
  and fourth order equations using legendre polynomials.
\newblock {\em Siam Journal of Scientific Coputing}, 15:1489--1505, 1994.

\bibitem{Shen1995}
J.~Shen.
\newblock Efficient spectral-galerkin method ii. direct solvers of second
  fourth order equations by using chebyshev polynomials.
\newblock {\em SIAM Journal of Scientific Computing}, 16(1):74--87, 1995.

\bibitem{SumerJensenSorensenFredsoeLiuCarstensen2010}
B.~M. Sumer, P.~M. Jensen, L.~B. S{\o}rensen, J.~Freds{\o}e, P.~L.-F. Liu, and
  S.~Carstensen.
\newblock Coherent structures in wave boundary layers. part 2. solitary motion.
\newblock {\em Journal of Fluid Mechanics}, 646:207--231, 2010.

\bibitem{TrefethenTrefethenReddyDriscoll1993}
L.~N. Trefethen, A.~E. Trefethen, S.~C. Reddy, and T.~A. Driscoll.
\newblock Hydrodynamic stability without eigenvalues.
\newblock {\em Science}, 261:578--584, 1993.

\bibitem{VentselKrauthammer2001}
E.~Ventsel and T.~Krauthammer.
\newblock {\em Thin Plates and Shells}.
\newblock Marcel Dekker, 2001.

\bibitem{VerschaevePedersenTropea2018}
J.~C.~G. Verschaeve, G.~K. Pedersen, and C.~Tropea.
\newblock Non-modal stability analysis of the boundary layer under solitary
  waves.
\newblock {\em Journal of Fluid Mechanics}, 836:740--772, February 2018.

\end{thebibliography}

\end{document}